\documentclass[12pt]{article} 
\usepackage[hyperfootnotes=false]{hyperref}
\usepackage{epsfig,pifont}
\usepackage{float}
\usepackage{dsfont}
\usepackage{comment}
\usepackage{bbold}
\usepackage{color}
\usepackage{hyperref}
\usepackage{enumitem}
\usepackage{graphicx}
\usepackage{placeins}
\usepackage{amssymb} 
\usepackage{multirow}
\usepackage{caption}
\usepackage{gensymb}
\usepackage{subcaption}
\usepackage{amsmath}
\usepackage{amssymb}
\usepackage{graphicx}
\usepackage{geometry}
\usepackage{cite}

\newlength{\abstractwidth}
\newcommand{\be}{\begin{equation}}
\newcommand{\bea}{\begin{eqnarray}}
\newcommand{\eea}{\end{eqnarray}}
\newcommand{\beq}{\begin{equation}}
\newcommand{\ee}{\end{equation}}

\def\simleq{\; \raise0.3ex\hbox{$<$\kern-0.75em
\raise-1.1ex\hbox{$\sim$}}\; }
\def\simgeq{\; \raise0.3ex\hbox{$>$\kern-0.75em
\raise-1.1ex\hbox{$\sim$}}\; }

\usepackage{tikz}
	\usetikzlibrary{decorations.pathreplacing}
    \usetikzlibrary{patterns}
    \usetikzlibrary{decorations.pathmorphing}
\usepackage{caption}

\newcommand{\id}{\mathds{1}}

\usepackage{amsthm}
\theoremstyle{plain}

\newtheorem*{prop*}{Conjecture A}

\makeatletter
\g@addto@macro\normalsize{%
  \setlength\abovedisplayskip{10pt}
  \setlength\belowdisplayskip{20pt}
  \setlength\abovedisplayshortskip{10pt}
  \setlength\belowdisplayshortskip{20pt}
}
\makeatother

\newcommand\To{\rule{0pt}{4.5ex}}       
\newcommand\Bo{\rule[-3.0ex]{0pt}{0pt}} 

\newcommand\BoTiny{\rule[-2ex]{0pt}{0pt}} 

\def\lsim{ \lower .75ex \hbox{$\sim$} \llap{\raise .27ex
\hbox{$<$}} }
\def\gsim{ \lower .75ex \hbox{$\sim$} \llap{\raise .27ex
\hbox{$>$}} }

\def\su2n{U($2^N$)}

\def\bi{\begin{itemize}}
\def\ei{\end{itemize}}

\makeatletter
\g@addto@macro\normalsize{%
  \setlength\abovedisplayskip{5pt}
  \setlength\belowdisplayskip{5pt}
  \setlength\abovedisplayshortskip{5pt}
  \setlength\belowdisplayshortskip{5pt}
}
\makeatother
\renewcommand{\title}[1]{\vbox{\center\LARGE{#1}}\vspace{5mm}}
\renewcommand{\author}[1]{\vbox{\center#1}\vspace{5mm}}
\newcommand{\address}[1]{\vbox{\center\em#1}}

\begin{document}

  \begin{titlepage}

\rightline{}
\bigskip
\bigskip\bigskip\bigskip\bigskip
\bigskip

\centerline{\large \bf {Polynomial Equivalence of Complexity Geometries}}

\bigskip
\begin{center}

\author{Adam R. Brown}

\address{Google Research (Blueshift), Mountain View, CA 94043, USA}

\address{Stanford Institute for Theoretical Physics and Department of Physics, \\
Stanford University, Stanford, CA 94305, USA}


\begin{abstract}

\noindent  This paper proves the polynomial equivalence of a broad class of definitions of quantum computational 
complexity.  We study right-invariant metrics on the unitary group---often called `complexity geometries' following the definition of quantum  complexity proposed by Nielsen---and delineate the equivalence class of metrics that have the same computational power as quantum circuits. 
Within this universality class, any unitary that can be reached in one metric can be approximated in any other metric in the  class with a slowdown that is at-worst polynomial in the length and number of qubits and inverse-polynomial in the permitted error.   We describe the equivalence classes for two different kinds of error we might  tolerate: Killing-distance    error, and operator-norm error. All metrics in both equivalence classes are shown to have exponential diameter; all metrics in the operator-norm equivalence class are also shown to give an alternative definition of the quantum complexity class BQP. 

My results extend those of Nielsen \emph{et al.},~who in 2006 proved that one particular metric is polynomially equivalent to quantum circuits. The Nielsen \emph{et al.}~metric is incredibly highly curved. I show that the greatly enlarged equivalence class established in this paper  also includes metrics that have modest curvature. I argue that the modest curvature makes these metrics more amenable to the  tools of differential geometry,  and therefore makes them more promising starting points for Nielsen's program of using differential geometry to prove complexity lowerbounds.

	     In a previous paper my collaborators and I---inspired by the UV/IR decoupling that happens in the phenomenon of renormalization---conjectured that high-dimensional metrics that look very different at short scales will often nevertheless give rise at long scales to the same emergent effective geometry.  The results of this paper provide evidence for those conjectures, since many complexity metrics that have radically different penalty factors and therefore radically different short-distance properties are shown to belong to the same long-distance equivalence class. 
  \end{abstract}
 
 \end{center}


\let\thefootnote\relax\footnotetext{email: {mr.adam.brown@gmail.com}, accepted by \href{https://quantum-journal.org}{\emph{Quantum}} on 28 Jan 2024}

\end{titlepage}

 \newpage

   \tableofcontents

\section{Introduction \& Summary}

In this paper we prove the polynomial equivalence of a broad class of definitions of quantum 
complexity. This class includes many different complexity geometries that may differ exponentially at short distances  but at longer distances are all shown to be polynomially equivalent; the class also includes the 
orthodox 
gate-counting 
definition 
that involves approximating unitaries by compiling circuits from gates and then counting the gates.

A complexity geometry is a right-invariant Riemannian metric on the unitary group, characterized by a penalty schedule (see Sec.~\ref{subsec:woulditreallybeplaguarismificopiedthisfromthelastpaper}). We will delineate the class of complexity geometries that give rise to polynomial equivalence. We will show that within this equivalence class any unitary $U \in \textrm{U}(2^N)$ that can be reached in one metric with a path of  length $L$ can be approximated in any other metric in the class with a path whose length is at-worst polynomial in $L$ and $N$ and inverse-polynomial in the permitted error.

In the context of computational complexity, the study of right-invariant metrics was pioneered by Nielsen and collaborators \cite{Nielsen1,Nielsen2,Nielsen3,Nielsen4,NielsenSingleQubit}. In 2006, Nielsen, Dowling, Gu, \& Doherty  \cite{Nielsen2} proved that for one particular complexity geometry---the so-called `cliff metric', which exponentially punishes motion in any tangent direction that has terms that touch more than two qubits at once (for a review and rigorous definition, see Sec.~\ref{subsec:woulditreallybeplaguarismificopiedthisfromthelastpaper})---this gives rise to a definition of complexity that is polynomially equivalent to the standard gate-counting definition 
\begin{equation}
\textrm{length of path in cliff metric to get close} \ \sim \  \textrm{number of gates to get close} .  
\end{equation}
In this paper we will extend this result to less draconian penalty schedules. We will describe sufficient conditions on a penalty schedule $\mathcal{I}(\sigma_I)$ such that 
\begin{equation}
\textrm{length of path in cliff metric to get close} \ \sim \ \textrm{length of path in $\mathcal{I}(\sigma_I)$ metric to get close} .  \label{eq:headlineresult}
\end{equation}
From the point of view of geometry, this will be interesting because it will prove the long-distance polynomial equivalence of different right-invariant metrics on the unitary group, even though these metrics differ exponentially at short scales, advancing the `effective geometry' program laid out in \cite{Brown:2021rmz}. From the point of view of complexity, this will be interesting because it will provide an alternative definition of the set of tasks that can be efficiently performed on a quantum computer, and this alternative definition has primitive operations that are substantially more permissive than either the gate definition or the cliff-metric definition. Furthermore, many of the  complexity geometries that are shown to be in the equivalence class are considerably easier to work with than the cliff metric---and to deploy the tools of differential geometry against \cite{Brown:2021euk}---because unlike the extremely highly curved cliff metric, these geometries have only modest curvature. 

\subsection{Summary of results} \label{subsec:summaryofresults}
Let's describe our headline results. We will write down sufficient conditions on the complexity geometry in order for the polynomial equivalence Eq.~\ref{eq:headlineresult} to hold. Our theorems will only apply to complexity geometries for which the penalty metric (defined in Sec.~\ref{subsec:woulditreallybeplaguarismificopiedthisfromthelastpaper}) is diagonal in the generalized Pauli basis $\sigma_I$ (also defined in Sec.~\ref{subsec:woulditreallybeplaguarismificopiedthisfromthelastpaper}). For such metrics the `penalty schedule'  (also, like all terms in this summary, defined in Sec.~\ref{subsec:woulditreallybeplaguarismificopiedthisfromthelastpaper}) assigns a `penalty factor' $\mathcal{I}(\sigma_I)$ to each of the $4^N$ basis directions $\sigma_I$ in the tangent space to U($2^N$); this is called a penalty factor because a larger value of $\mathcal{I} (\sigma_I)$ makes the direction $\sigma_I$ more expensive to evolve in. 
In order to prove the polynomial equivalence, we will need to prove that the complexity geometries can efficiently emulate gates, and that gates can efficiently emulate the complexity geometries. 
\subsubsection{Complexity geometries that can efficiently emulate gates} 
The first direction of the equivalence will be straightforward. In order that the complexity geometries are guaranteed to be able to efficiently emulate gates, a sufficient condition is 
\begin{eqnarray}
& \textrm{\emph{for 2-local basis directions $\sigma_I$ (i.e.~basis directions that act non-trivially} } &  \label{sufficientcondition0} \\ 
&\textrm{ \emph{on only 2 qubits) the penalty factor $\mathcal{I}(\sigma_I)$ is at-most polynomially big}} &\nonumber 
\end{eqnarray} 
(We will show in Sec.~\ref{subsubsec:imcrementalimprovements} that this condition can be significantly relaxed.) 
\subsubsection{Complexity geometries that can be efficiently emulated by gates} 
The other direction of the equivalence will require more work.  In order that gates are guaranteed to be able to efficiently emulate the complexity geometries, we will need the penalty schedule to be sufficiently punitive, meaning it assigns a small value of $\mathcal{I}$ to only a small number of basis directions $\sigma_I$. To make this precise, we will have to be precise about what we mean in Eq.~\ref{eq:headlineresult}
by `close'. We will consider two different definitions of `close', leading to different sets of sufficient conditions. 
\begin{enumerate}
\item Killing close. 
If we demand that we get polynomially close as measured in the Killing metric (defined in Sec.~\ref{subsec:defofbiinvariantriemannian}), the sufficient condition for polynomial equivalence is 
\begin{eqnarray}
& \textrm{\emph{for any polynomially large $\overline{\mathcal{I}}$,  there are only}} &  \label{sufficientcondition1} \\ &\textrm{ \emph{polynomially many basis directions $\sigma_I$ with $\mathcal{I}(\sigma_I) < \overline{\mathcal{I}}$}} &\nonumber 
\end{eqnarray} 
\item Operator-norm close. If we make the more demanding demand that we get polynomially close as measured by the operator-norm distance (defined in Sec.~\ref{subsec:definitionofbiinvariantmatrixnorms}) then a sufficient condition for polynomial equivalence is 
\begin{eqnarray}
& \textrm{\emph{  for any polynomially small value, we can make the harmonic sum }} &\nonumber \\ &\textrm{ \emph{of the penalty factors $\sum_{\sigma_I} \mathcal{I}(\sigma_I)^{-1}$ smaller than that value by    }} &  \label{sufficientcondition2} \\
& \textrm{ \emph{omitting from the sum at most polynomially many  of the $\sigma_I$s} } & \nonumber 
\end{eqnarray} 
Another sufficient condition for operator-norm polynomial equivalence is 
\begin{eqnarray} 
& \textrm{\emph{for any polynomially large $\overline{\mathcal{I}}$,  there are only}} &  \label{sufficientcondition3} \\ &\textrm{ \emph{polynomially many basis directions $\sigma_I$ with $\mathcal{I}(\sigma_I) < 2^N \overline{\mathcal{I}}$}} &\nonumber 
\end{eqnarray} 
\end{enumerate}  
\noindent We evaluate these sufficient conditions against some example penalty schedules in Table~\ref{table:summaryofresults}. \\

\noindent When fed the same input state, two unitaries that are close in operator norm will give output states that are close in inner-product. A corollary of our results therefore is that, given the power to implement any polynomial-length path through a complexity geometry  that satisfies conditions  \ref{sufficientcondition0} and \ref{sufficientcondition2} or \ref{sufficientcondition3}, the set of decision problems that can be solved with high probability for every input is exactly the quantum complexity class 
 `BQP'.

 \begin{center}
  \begin{table}
   \begin{center}
\begin{tabular}{|c||c|c|}  
\hline 
& polynomial $\mathcal{C}_\textrm{gates}$ can & polynomial $\mathcal{C}_\textrm{gates}$ can \To  \\ 
&  get Killing close
 &  get $||\cdot||_\textrm{op}$ close   \Bo \\
\hline 
\hline 
bi-invariant &   &  \To \\
$\mathcal{I}_{k} = 1$ &   &    \\
\textrm{(Killing metric)} & \hspace{4cm}  &   \hspace{4cm}  \Bo \\
\hline 
hard cliff &   &  \To  \\
$\mathcal{I}_{k \leq 2} = 1 $, $\mathcal{I}_{k \geq 3} = \alpha^N $  & \ding{51}  &\ding{51} \\
for some $\alpha > \sqrt{2}$&   &   \Bo \\
\hline 
soft cliff &   & \To \\
$\mathcal{I}_{k \leq 2} = 1 $,  $\alpha^N < \mathcal{I}_{k \geq 3} \leq 2^{ \frac{1}{2} N} $ &   \ding{51} &  \\
for some  $\alpha >1$ &   &  \Bo \\
\hline 
binomial &   &  \To \\
$\mathcal{I}_{k } = (\mathcal{N}_k)^\alpha \equiv ( {N \choose k} 3^k )^\alpha  $ & \ding{51}  &\ding{51}  \\
for some  $\alpha >1$ &   & \Bo \\
\hline  
binomial &   &  \To \\
$\mathcal{I}_{k } = (\mathcal{N}_k)^\alpha \equiv ( {N \choose k} 3^k )^\alpha  $&   \ding{51} &  \\
for some  $0<\alpha \leq1$ &   &  \Bo \\
\hline  
exponential &   &  \To \\
$\mathcal{I}_{k } =x^{2k}  $&   \ding{51} &  \\
for some  $x >1$& (quasi-polynomial)  &  \Bo \\
\hline  
 \end{tabular}
\caption{Summary of results for some example penalty schedules. For all these examples, the penalty schedule $\mathcal{I}(\sigma_I)$ is a function only of the weight $k$ of the direction $\sigma_I$, see Sec.~\ref{subsec:complexitygeometrydefinition} (though our theorems are more general). 
A \ding{51}  means that if there exists a polynomial-length path in that penalty metric, there must also exist a circuit that approximates it with a number of gates that is polynomial in the length, the number of qubits, and the targeted error. For the first column, the error is measured in the Killing metric (or equivalently in the normalized Frobenius norm $||\textrm{error}||_{\overline{F}}$, see Sec.~\ref{subsec:definitionofbiinvariantmatrixnorms}); for the second column, the error is measured in the operator norm $||\textrm{error}||_\textrm{op}$. 
} \label{table:summaryofresults}
\end{center}
\end{table}
\end{center}

\section{Defining distance functions on the unitary group} \label{subsec:woulditreallybeplaguarismificopiedthisfromthelastpaper}

In this paper we will consider distances functions on U($2^N$), the group of purity-preserving linear functions on $N$ qubits. As well as satisfying all the required axioms---symmetry, the triangle inequality, etc.---all these distance functions will in addition be `right invariant', meaning that 
\begin{equation}
\textrm{right-invariance:} \ \ \ \ \textrm{distance}(U_1,U_2) = \textrm{distance}(U_1 U_R,U_2 U_R) \textrm{ for any unitary $U_R$}. 
\end{equation} 
Let's define the distance functions now. 

\subsection{Killing metric (bi-invariant Riemannian)} \label{subsec:defofbiinvariantriemannian}
The  simplest and most symmetrical nontrivial Riemannian distance function on the unitary group is the Killing metric. The infinitesimal distance between two nearby unitaries $U$ and $U + dU$ is defined as 
\begin{equation}
ds^2 =  \overline{\textrm{Tr}} [dU^\dagger dU ]=  \sum_{I,J} (i \overline{\textrm{Tr}}  \, \sigma_I  U dU^\dagger) \delta_{IJ}(i \overline{\textrm{Tr}}  \, \sigma_J  U dU^\dagger)    \ . \label{eq:biinvariantmetric}
\end{equation}
Let's explain what these terms mean. First the bar over the $\overline{\textrm{Tr}}$ indicates we have normalized the trace so that it gives the average---not the sum---of the diagonal terms, 
\begin{equation}
 \overline{\textrm{Tr}}  [\mathds{1}] = 1 \ ; \label{eq:normalizationoftrace}
\end{equation}  
when applied to a $2^N \times 2^N$ matrix this means $\overline{\textrm{Tr}}  = 2^{-N} {\textrm{Tr}}$. 
The symbol $\delta_{IJ}$ is the Kronecker delta. The sum runs over all $4^N$ of the generalized Pauli's $\sigma_I$,  which are defined as  
\begin{equation} 
\sigma_I \in 
\left\{ \begin{array}{c}
\mathds{1} \\
\sigma_x \\
\sigma_y \\
\sigma_z
\end{array}
\right\}_1 
\otimes \, 
\left\{ \begin{array}{c}
\mathds{1} \\
\sigma_x \\
\sigma_y \\
\sigma_z
\end{array}
\right\}_2 
\otimes \, \ldots \, \otimes  \, 
\left\{ \begin{array}{c}
\mathds{1} \\
\sigma_x \\
\sigma_y \\
\sigma_z
\end{array}
\right\}_N . \label{eq:tensorfactors}
\end{equation}
The generalized Pauli's  give a complete basis for the tangent space of the unitary group; this basis is orthonormal since $\overline{\textrm{Tr}}[\sigma_I \sigma_J] = \delta_{IJ}$. 
The `weight' or `$k$-locality' of a generalized Pauli $\sigma_I$ is defined as the number of qubits on which it acts nontrivially. In other words, a $k$-local $\sigma_I$ picks $k$ of the terms in the tensor product Eq.~\ref{eq:tensorfactors} to be SU(2) Pauli matrices $(\sigma_x,\sigma_y,$ or $\sigma_z)$, and $N-k$ terms to be $\mathds{1}$. The number of exactly $k$-local generalized Pauli's is 
\begin{equation}
\mathcal{N}_k \equiv {N \choose k} 3^k \ . 
\end{equation}
 Using the generalized Pauli's, we can  decompose the instantaneous tangent Hamiltonian, which Schr\"{o}dinger's equation $-i \dot{U} = HU$ tells us is given by $H \equiv -i \dot{U} U^\dagger$, as 
\begin{equation}
H(t) = \sum_I h_I(t) {\sigma}_I \ . \label{eq:canonicalHamiltonian}
\end{equation}
Eq.~\ref{eq:biinvariantmetric} implies that the length of a small step that applies this Hamiltonian for a time $\delta$ is 
\begin{equation}
s = \sqrt{ \overline{\textrm{Tr}} H^2(t)} \delta = \sqrt{\sum_I h_I^2(t)} \delta \ . \label{eq:TrHsquared}
\end{equation}
So far we have defined the length of infinitesimal paths. The length of a general differentiable path  through the space of unitaries is defined by breaking the path up into infinitesimal segments and summing the lengths of the segments. Finally the Killing distance $s(U_1,U_2)$ between two unitaries is given by the length of the shortest path that connects them. As discussed in Appendix~\ref{sec:Frobboundsinnerproduct}, this distance  is never be bigger than $\pi$, which is the distance between two antipodal unitaries (e.g.~$\mathds{1}$ and $- \mathds{1}$). \\

\noindent The Killing distance function is  `bi-invariant', meaning it is invariant under both left-multiplication and right-multiplication, 
\begin{equation}
s(U_1,U_2) = s(U_L U_1 U_R, U_L U_2 U_R) \ \ \textrm{ for all unitaries } U_L \ \& \  U_R \ . 
\end{equation}
One consequence of bi-invariance is that  errors at-worst add, 
 \begin{equation}
s(U_1 U_2,U_{1'} U_{2'}) \leq  s(U_1 U_2, U_1 U_{2'}) + s(U_1 U_{2'}, U_{1'} U_{2'}) =  s(U_2 , U_{2'} ) + s( U_1, U_{1'}) \ . \label{eq:biinvariantriemanniancompose}
 \end{equation}
(Here the first step uses the triangle inequality and the second step uses the bi-invariance.) Applying this equation to the unitaries built from two different Hamiltonians gives 
\begin{equation}
s(\mathcal{P} e^{i \int H_1(t) dt} ,  \mathcal{P} e^{i \int H_2(t) dt} ) \ \leq \  \int \sqrt{ \overline{\textrm{Tr}} [ (H_1(t) - H_2(t) )^2} ] dt \ ,  \label{eq:distancebetweentwodifferentpaths}
\end{equation}
where ``$\mathcal{P}$'' is Dyson's path-ordering operator, necessary because the Hamiltonian at one time may not commute with the Hamiltonian at a different time.

\subsection{Complexity geometry} \label{subsec:complexitygeometrydefinition}
The complexity metrics generalize the Killing metric by replacing $\delta_{IJ}$ with a  positive symmetric `penalty matrix' $\mathcal{I}_{IJ}$, so that the line-element becomes \begin{equation}
dL^2 =  \sum_{I,J} (i \overline{\textrm{Tr}} \, \sigma_I  U dU^\dagger) \mathcal{I}_{IJ}(i \overline{\textrm{Tr}} \, \sigma_J  U dU^\dagger) \ . \label{eq:rightinvariantmetric}
\end{equation}
(For a pedagogical introduction see \cite{Brown:2019whu}; for other recent work see \cite{qutrit,Brown:2016wib,secondlaw,Lin:2018cbk,Balasubramanian:2019wgd,Bhattacharyya:2019kvj,Caginalp:2020tzw,Auzzi:2020idm,Lin:2019kpf,Yan:2020twr,Balasubramanian:2021mxo,Bulchandani:2021yov,Wu:2021pzg,Basteiro:2021ene,Chapman:2017rqy,Jefferson:2017sdb,Hackl:2018ptj,Bhattacharyya:2018bbv,Khan:2018rzm}) 
This metric assigns a length to infinitesimal paths through the space of unitaries; the length of a non-infinitesimal path is the sum of the lengths of its segments; and the distance between two unitaries is the length of the shortest path that connects them. This expression is manifestly still right-invariant but is in general not left-invariant. 
In this paper, we will prove theorems only about metrics for which the penalty matrix $\mathcal{I}_{IJ}$ is diagonal in the generalized-Pauli basis, 
\begin{equation}
\textrm{diagonal penalty matrix: } \ \ \  \ \ \ \  \mathcal{I}_{IJ} = \delta_{IJ}  \mathcal{I}(\sigma_I) \ .  \hspace{3cm} \ \    \ \ \ \ \ \ \ \label{eq:diagonalpenaltymatrix}
\end{equation} 
(We discuss the necessity of this condition in Sec.~\ref{subsubsec:imcrementalimprovements}.) The diagonal component $\mathcal{I}(\sigma_I)$ is known as the `penalty factor' since it stretches the tangent direction $\sigma_I$ so that motion in  that direction more expensive (or cheaper for $\mathcal{I}(\sigma_I) <1$). The length assigned to the small step  generated by applying the Hamiltonian of Eq.~\ref{eq:canonicalHamiltonian} for an infinitesimal time $\delta$ is 
\begin{equation}
L = \sqrt{\sum_I \mathcal{I}(\sigma_I) h_I^2} \, \delta \ . 
\end{equation}

\noindent  It follows directly from the metric Eq.~\ref{eq:rightinvariantmetric}  that when one penalty schedule is harder than another for every $\sigma_I$, the harder schedule  always gives longer distances
 \begin{equation}
\mathcal{I}(\sigma_I) \ \leq \ \tilde{\mathcal{I}}(\sigma_I)  \ \longrightarrow \  L (U_1,U_2) \ \leq \  \tilde{L}(U_1,U_2) \ . 
\end{equation}
Applying this formula to compare a complexity geometry with largest penalty factor $\mathcal{I}_\textrm{max}$ to the scaled Killing metric that has $\mathcal{I}(\sigma_I) = \mathcal{I}_\textrm{max}$ in every direction tells us that 
\begin{equation}
L(U_1,U_2) \ \leq \ \pi \sqrt{ \mathcal{I}_\textrm{max}} \ . \label{eq:lessthanpiImax}
\end{equation} 

\vspace{5mm} 
\noindent {\bf Example metrics}. The theorems we will prove will apply to all penalty metrics that are diagonal in the generalized Pauli basis (i.e.~that have the form in Eq.~\ref{eq:diagonalpenaltymatrix}), but we will illustrate the theorems with three  metrics of special interest. For all three, the penalty factor $\mathcal{I}(\sigma_I)$ assigned to a given generalized Pauli is a function only of the direction's $k$-locality. We can completely specify such penalty metrics by specifying the value of the penalty $\mathcal{I}_k$ for each of the possible values of the weight $k$ (i.e.~for every whole number between 0 and $N$). The three families of special metrics are characterized by the parameters $\mathcal{I}_\textrm{cliff}$, $\alpha$ and $x$: 
\begin{eqnarray}
\textrm{`cliff metric' } \ \ \ \ \ \ \ \  \ \ \ \ \ \  \mathcal{I}_k & = & \biggl\{ \begin{array}{lll}
1 & \textrm{ if } & k \leq 2  \\   \mathcal{I}_\textrm{cliff} & \textrm{ if } & k \geq 3 
 \end{array} \label{eq:examplemetric1} \\
\textrm{`binomial metric' } \ \ \ \ \ \   \ \  \mathcal{I}_k  &= & \left(  \mathcal{N}_k\right)^\alpha  \equiv \left(  {N \choose k} 3^{k} \right)^\alpha \Bo \label{eq:examplemetric2}  \\
\textrm{`exponential metric' } \ \  \ \    \mathcal{I}_k  &=& x^{2k } \  \Bo . \label{eq:examplemetric3} 
\end{eqnarray}
For all three metrics, 1- and 2-local directions are cheap, and large-$k$ directions are expensive; the metrics differ in how gradually the cost increases as a function of $k$. The cliff metric was the original  metric considered in \cite{Nielsen1,Nielsen2}. The exponential metric was discussed \cite{secondlaw}, where it was pointed out that its modest sectional curvature makes it a good candidate to reproduce the correct `switchback effect' seen in quantum chaos and holographic black holes. The binomial metric also has modest sectional curvature, and was discussed in \cite{Brown:2021rmz} as a possible candidate for the `critical metric' to which all harder metrics flow in the infrared.

\subsection{Gate complexity} \label{subsec:gatecomplexity}

The gate distance between two unitaries is the number of gates in the smallest circuit that connects them 
\cite{Nielsen:2011:QCQ:1972505}. In the version of gate complexity we will consider in this paper, we will take the primitive operations to be the action of general two-qubit gates. At each step we are allowed to pick any two qubits, and act on those two qubits with any two-qubit unitary, i.e.~any element of U($4$). In this way we can build up any element of U($2^N)$ exactly with no more than about $4^N$ gates \cite{Knill:1995kz}, 
\begin{equation}
\forall U_1 \forall U_2 , \ \ \ \ \ \mathcal{C}_\textrm{gates}(U_1,U_2)  < N^2 4^N \ . \label{eq:trivialbound}
\end{equation}
$\mathcal{C}_\textrm{gates}(U_1,U_2)$ is defined as the number of gates in the smallest circuit that implements $U_2^\dagger U_1$.  \\

\noindent It follows directly from the definitions that the cliff metric gives a distance function that is smaller than ($\pi$ times) the number of gates in the smallest circuit, 
\begin{equation}
L_\textrm{cliff}(U_1,U_2) \leq \pi \, \mathcal{C}_\textrm{gates}(U_1,U_2) \ . \label{eq:clifflessthanpigates}
\end{equation}
This is because any element of U(4) can be generated with a time-independent $\overline{\textrm{Tr}} H^2 = 1$ Hamiltonian acting only on those 2 qubits for a time at most $\pi$. \\

\noindent (In this paper, we have chosen a definition of gate complexity in which there is a continuum of primitive operations---any element of U(4) is an allowed gate. However, this set of gates is known to be polynomially equivalent---polynomial in both the complexity and in the targeted error---to a more restrictive definition of complexity that allows only a discrete set of primitive gates, such as CNOT plus Hadamard plus a random single-qubit phase. This is because of the Solovay-Kitaev theorem \cite{SolovayKitaev1,SolovayKitaev2}.)

 \begin{center}
 \begin{table}
\begin{tabular}{c||c|c|c|c|c|} 
&triangle & right- & left-   & continuous  & Riemannian \\
& inequality   &    invariant   &   invariant   &  &    \\
  \hline
  \hline
Killing distance  & \ding{51} & \ding{51} & \ding{51} & \ding{51} &  \ding{51} \To \Bo \\
\hline
complexity geometry  & \ding{51}& \ding{51}& & \ding{51}& \ding{51} \To \Bo\\
\hline
gate complexity  & \ding{51}&\ding{51}& && \To \Bo \\
\hline
Frobenius norm  & \ding{51}&\ding{51} &\ding{51}&\ding{51}& \To \Bo\\
\hline
operator norm  & \ding{51}&\ding{51}&\ding{51}&\ding{51} & \To \Bo\\
\hline
\end{tabular}
\caption{Properties of the five kinds of distance functions we consider in this paper.} \label{table:summaryofproperties}
\end{table}
\end{center}

\subsection{Bi-invariant matrix norms} \label{subsec:definitionofbiinvariantmatrixnorms}
We can also define a distance between two unitaries via a matrix norm. The distance function is defined as  the norm of the difference of the two unitaries
\begin{equation}
\textrm{distance}_\textrm{norm}(U_1,U_2) \equiv ||U_1 - U_2||_\textrm{norm}  \ . 
\end{equation}
The matrix norms we will consider will all be bi-invariant, which means 
\begin{equation}
||U_1 -  U_2|| =  || U_L U_1 U_R - U_L U_2 U_R|| \ \ \textrm{ for all unitaries } U_L \ \& \  U_R \ .
\end{equation}
Bi-invariant distance functions have the important  property that errors compose (as we already saw with the Killing distance in Eq.~\ref{eq:biinvariantriemanniancompose}), 
\begin{equation}
|| U_1 U_2 - U_{1'} U_{2'} || \leq || U_1 - U_{1'} || + || U_2 - U_{2'} ||  \ . \label{eq:compositionoferrorsfornorms}
\end{equation}
This follows from combining the triangle inequality $|| U_1 U_{2} - U_{1'} U_{2'} || = || U_1 U_{2}  - U_1 U_{2'}  + U_1 U_{2'} -  U_{1'} U_{2'} || \leq || U_1 U_{2}  - U_1 U_{2'}|| + || U_1 U_{2'} -  U_{1'} U_{2'}  ||$ with bi-invariance $|| U_1 U_{2}  - U_1 U_{2'}|| + || U_1 U_{2'} -  U_{1'} U_{2'}  || = || U_2  - U_{2'} || + ||  U_1 -  U_{1'} ||$.  For unitaries generated by Hamiltonians, it follows that for any bi-invariant norm 
\begin{equation}
|| \mathcal{P} e^{i \int H_1(t) dt}  -  \mathcal{P} e^{i \int H_2(t) dt} || \leq \int dt || H_1(t) - H_2(t)||  \ .  \label{eq:normpathlessthanhamiltonian} 
\end{equation}

\subsubsection{Frobenius norm} \label{subsec:frobeniusnormdefinition}
 The Frobenius norm of a matrix $A$ is defined as 
\begin{equation}
||A||_{F} \equiv \sqrt{{\textrm{Tr}}[A^\dagger A] } \ . 
\end{equation}Because of the cyclic property of the trace, the Frobenius distance is bi-invariant.  
It will also be helpful to define the normalized Frobenius norm 
\begin{equation}
||A||_{\overline F} \equiv \sqrt{\overline{\textrm{Tr}}[A^\dagger A] } \ . 
\end{equation}
For U$(2^N)$ the normalized trace is given by Eq.~\ref{eq:normalizationoftrace} as $\overline{\textrm{Tr}} = 2^{-N}\textrm{Tr}$ so the norms are related by $||A||_{\overline{F}} =  2^{-N/2}||A||_{{F}}$. The Frobenius norm is the squareroot of the \emph{sum} of eigenvalues of $A^\dagger A$, whereas the normalized Frobenius norm is the squareroot of the \emph{average}. \\

\noindent The normalized Frobenius norm of the Hamiltonian of Eq.~\ref{eq:canonicalHamiltonian} is given by 
\begin{equation}
||H||_{\overline{F}} = ||\sum_I h_I  {\sigma}_I ||_{\overline{F}} = \sqrt{ \overline{\textrm{Tr}}{H^2}} = \sqrt{\sum_I h_I^2}  \ . \label{eq:FrobeniusnormofHamiltonian}
\end{equation}
The normalized-Frobenius-norm length of the \emph{infinitesimal} path generated by $H$ is 
\begin{equation}
||\mathds{1} - e^{i H \delta}||_{\overline{F}} = ||H||_{\overline{F}} \delta + O(\delta^2) = \sqrt{ \overline{\textrm{Tr}}H^2} \delta + \textrm{O}(\delta^2) \ . 
\end{equation}
Comparing with Eq.~\ref{eq:TrHsquared}, we recognize this as being the same as the Killing length 
\begin{equation}
||\mathds{1} - e^{i H \delta}||_{ \overline{F}} \ = \ s(\mathds{1} , e^{i H \delta}) - O(\delta^2) \ . 
\end{equation} 
At longer separations, Eq.~\ref{eq:normpathlessthanhamiltonian} tells us that the normalized Frobenius distance is no greater than (and is typically less than) the Killing distance
\begin{equation}
||U_1 - U_2||_{ \overline{F}} \ \leq \ s(U_1,U_2) \ . \label{eq:RiemannianbiggerthanFrobenius}
\end{equation} 
Both distance functions satisfy the triangle inequality, but only for the Killing distance is there always an intermediate point that saturates the triangle inequality. The Frobenius-norm distance between $U_1$ and $U_2$ is typically shorter than the sums of the lengths of the path-segments for every connecting path, and is therefore not Riemannian.

\begin{figure}[htbp] 
    \centering
    \includegraphics[width=2.4in]{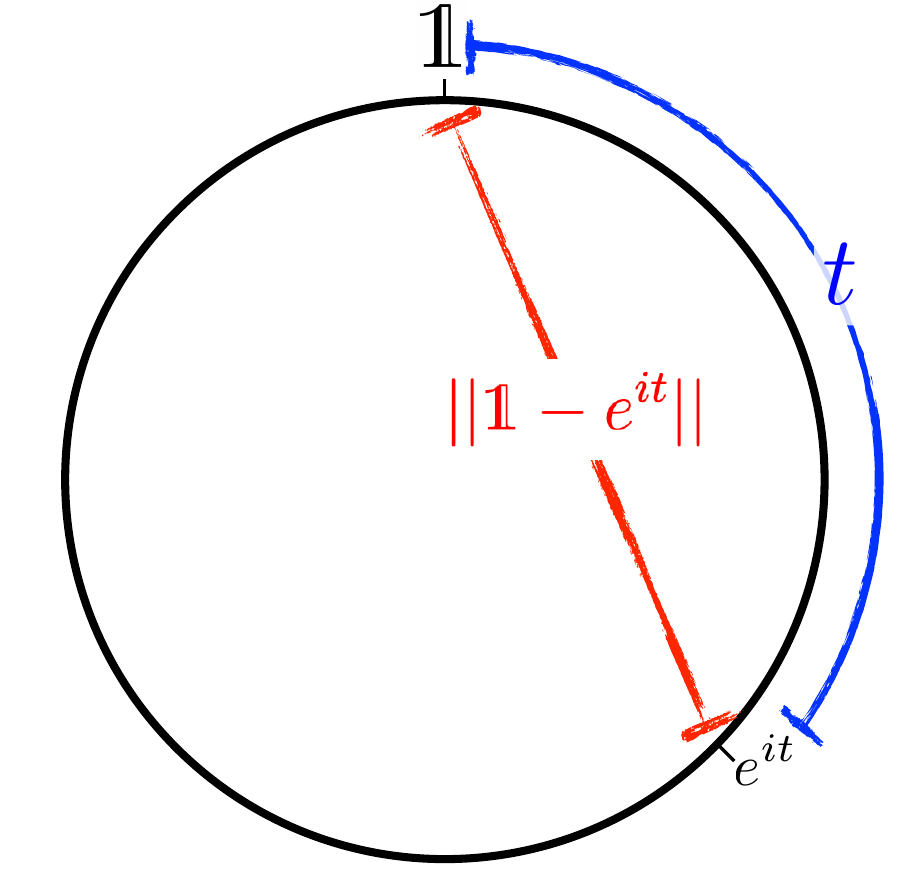} 
    \caption{The group U(1) is a circle. The \textcolor{blue}{Killing distance} from the identity to $U = e^{it}$ is the distance  around this circle, $\color{blue}{s(\mathds{1} , e^{it}) = t}$. The \textcolor{red}{normalized Frobenius-norm distance} is the length of the chord, $\color{red}{||\mathds{1} - e^{it}||_{\overline{F}} = 2 \sin t/2}$.}
        \label{fig:chordorcircumference}
 \end{figure}

The relationship between the two distance functions can be understood by considering the simplest  case: U(1). As depicted in Fig.~\ref{fig:chordorcircumference}, the group U(1) is a circle. The Killing distance between two unitaries is the length around the circumference of this circle, whereas the Frobenius distance is the length of the chord through the circle. These two distances agree for infinitesimal separations; for non-infinitesimal separations the Frobenius-norm distance is shorter.
\FloatBarrier 
\noindent While the Killing distance $s$ is longer than the normalized Frobenius-norm distance, we will prove in Appendix~\ref{sec:Frobboundsinnerproduct} that it's never that much longer, 
\begin{equation}
 ||U_1 - U_2||_{ \overline{F}} \  \leq \  s(U_1,U_2) \ \leq \  \frac{\pi}{2} ||U_1 - U_2||_{ \overline{F}} \ .  \label{eq:tobeprovedinappendix1}
\end{equation} 
Looking at the U(1) case in  Fig.~\ref{fig:chordorcircumference}, we see that the lowerbound is saturated for neighboring unitaries and the upperbound is saturated for antipodal unitaries. What Eq.~\ref{eq:tobeprovedinappendix1} means in practice is that we can bound the Killing distance by bounding the normalized Frobenius-norm distance, which will be useful since the Frobenius norm is often easier to work with.

\subsubsection{Operator norm} \label{subsubsec:operatornormerror}
The operator norm of a matrix $A$ is defined by 
\begin{equation}
||A||_\textrm{op} \equiv \max_{|\psi \rangle}  \sqrt{ \frac{   \langle \psi | A^\dagger A | \psi \rangle  }{    \langle \psi   | \psi \rangle } }  \ . \label{eq:operatornormdefinition}
\end{equation}
The operator norm is the squareroot of the \emph{largest} eigenvalue of $A^\dagger A$; two unitaries can only be operator norm close if they have close to the same effect on every state, including the worst-case state. It follows immediately from their definitions that for any matrix $A$, 
\begin{equation}
\hspace{1cm} ||A||_{ \overline{F}} \ \leq \ ||A||_{ \textrm{op}} \ \leq \ ||A||_{  F}  \equiv 2^{N/2} ||A||_{ \overline{F}} \ . \label{eq:norminequalitieswith2totheN}
\end{equation}
We can write a possibly tighter upperbound on $||H ||_\textrm{op}$ when $H$ is a Hamiltonian that only has support on at most $\mathcal{N}$ of the generalized Pauli's. The subadditivity of the norm tells us that 
\begin{equation}
||H||_\textrm{op} =  ||\sum_I^\mathcal{N} h_I \sigma_I ||_\textrm{op} \ \leq \ \sum_I^\mathcal{N} || h_I  \sigma_I ||_\textrm{op} = \sum_I^\mathcal{N} | h_I | \ . \label{hoplessthansumh}
\end{equation}
This inequality will be saturated if, for example, the only nonzero generalized Pauli's are those drawn from the $2^N$ operators composed exclusively of $\mathds{1}$ and $\sigma_z$ operators, but the inequality is typically \emph{not} saturated and is never saturated for $\mathcal{N} > 2^N$. Combining Eq.~\ref{hoplessthansumh} with the Cauchy-Schwarz inequality and with the definition Eq.~\ref{eq:FrobeniusnormofHamiltonian} gives  
\begin{equation}
||H||_\textrm{op} \ \leq \ \sum_I^{\mathcal{N}} |h_I| \ \leq \ \sqrt{ \mathcal{N}  \sum_I^{\mathcal{N}} h_I^2 } =  \sqrt{\mathcal{N}} ||H||_{\overline{F}} . \label{eq:norminequalitieswithN}
\end{equation}
In summary, our two bounds Eqs.~\ref{eq:norminequalitieswith2totheN} and \ref{eq:norminequalitieswithN} together give 
\begin{equation}
||H||_{ \overline{F}} \ \leq \ || H ||_\textrm{op} \ \leq \ \textrm{min}[\sqrt{\mathcal{N}},2^{N/2}] \  || H ||_{ \overline{F} } \ .  \label{eq:norminequalitieswithboth} 
\end{equation}
 \begin{table}
  \begin{center}
\begin{tabular}{|c||c|c|c|} 
\hline 
&\ \ \ \ \  \ $||H||_{\overline{F}} $ \ \   \ \ \ \  & \  \ \  \ \ $ ||H ||_\textrm{op} $\  \ \ \ \ & \ \ \ \ \ \ $||H||_{{F}} $ \ \  \ \ \   \To \Bo \\
\hline \hline 
& & & \\
$H $ & $\sqrt{ \vphantom{ \overline{F} )} \overline{ \textrm{Tr}} H^2 }  $ & $ \sqrt{ \vphantom{ {\overline{\overline{F}} } )} \textrm{max}_{|\psi \rangle}  \frac{ \langle \psi | H^2 | \psi \rangle}{\langle \psi | \psi \rangle} } $ & $\sqrt{ \vphantom{ \overline{F} )} {\textrm{Tr}} H^2 }    $\\
& & & \\
\hline 
$H = \sum_{I \in \{ \mathds{1}, \sigma_z \}^{\otimes N}  } h_I \sigma_I $ & $\sqrt{  \vphantom{ \overline{F} )} \sum_I h_I^2 }     $&  $\sum_I |h_I|    $ & $2^{N/2}\sqrt{ \vphantom{ \overline{F} )}\sum_I h_I^2 }     $  \To \Bo  \\
\hline
$h_1 = 1$ &&& \To \\ 
$h_{I \neq 1} = 0$  &$  1   $&$  1    $ & $2^{N/2}   $ \BoTiny  \\
$H =  \textrm{diag}[1,-1,1,-1,\ldots,1,-1]  $ & & &\Bo \\
\hline
$h_I = 2^{-N/2} \textrm{ if only $\mathds{1}$ or $\sigma_z$} $ &&& \To \\ 
$h_{I} = 0 \textrm{ if  contains  $\sigma_x$ or $\sigma_y$} $  &$   1  $&$ 2^{N/2}      $ & $2^{N/2}  $\BoTiny \\
$H = \textrm{diag}[1,0,0,0,\ldots,0,0] $ &  & & \Bo \\
\hline
\end{tabular}
\caption{The operator norm $||H||_\textrm{op}$ is always between $||H||_{\overline{F}}$ and $||H||_F$.
 The operator norm will be at the bottom end of this range when the eigenvalues of $H$ all have equal magnitude, for example any single generalized Pauli such as 
$H = \sigma_z \otimes \mathds{1} \otimes \sigma_x \otimes \ldots \otimes \sigma_x$.  By contrast the operator norm will be at the top end of this range when the eigenvalues are very unequal, so that $H$ changes a few states a lot but most states only a little or not at all, for example the \emph{control-control-control-\ldots-control-control-}$(\mathds{1}+\sigma_z)$ Hamiltonian is an equal superposition over $2^N$ generalized Pauli's. Note that for a general superposition of generalized Pauli's,  $||H||_\textrm{op}$ will typically be strictly less than $\sum_I |h_I |$ .} \label{Table:comparingoperatornormtofrobenius}
\end{center}
\end{table}

\FloatBarrier

\section{Approximating geodesics with gates} \label{sec:sec3}

In this section we will consider paths of length $L$ through a complexity geometry with arbitrary  penalty schedule $\mathcal{I}(\sigma_I)$. We will show how to approximate such paths within a given error tolerance using 2-local gates, and upperbound the number $\mathcal{C}_\textrm{gates}$ of 2-local gates required. For a broad class of penalty schedules, we will show that the required number of gates is at most polynomial in $L$ and the error. This generalizes the result of Nielsen, Dowling, Gu, \& Doherty  \cite{Nielsen2}, who proved the polynomial equivalence of the gate definition of complexity and the complexity geometry for the special case of the `cliff schedule' $\mathcal{I}_{1} = \mathcal{I}_2 = 1$, $\mathcal{I}_{k \geq 3} \geq 4^N$. 
Our general proof strategy will roughly follow that of Ref.~\cite{Nielsen2}, albeit having to contend with a number of complications that arise from the fact that the penalty schedules we are considering may assign exponentially smaller penalty factors than those assigned by the cliff schedule. \\

\noindent Consider the minimal path through the complexity geometry that constructs $U_\textrm{target}$. This path will be characterized by a (typically time-dependent) Hamiltonian $H(t)$
\begin{equation}
U_\textrm{target} = \mathcal{P}  \exp\Bigl[ i \int  {H}(t) dt \Bigl]  = \mathcal{P}  \exp\Bigl[ i \int \sum_I h_I(t)  {\sigma}_I dt \Bigl] \ .
\end{equation}
Without loss of generality we can normalize the time so that $H(t)$ has unit trace 
\begin{equation}
\overline{\textrm{Tr}}H(t)^2 = \sum_I h_I(t)^2 = 1  \ .  \label{eq:tracenormalization} 
\end{equation} 
Since the minimal path is a geodesic, the `difficulty' $\Gamma$ of this path, 
\begin{equation}
\Gamma^2 \equiv \langle H(t),H(t) \rangle \equiv \sum_I \mathcal{I}(\sigma_I) h_I(t)^2 \ , \label{eq:conservedhardness}
\end{equation}
will be a constant of motion\footnote{One way to see that $\Gamma$ is conserved is to plug $K = H$ into the Arnold-Khesin equation, which says that for any operator direction $K$, the equation of motion is 
\begin{equation}
\langle \dot{H},K \rangle = i \langle H, [H,K] \rangle \  . \label{eq:ArnoldKhesin}
\end{equation}
The conservation of $\Gamma$ is the analog of the conservation of total angular momentum for a spinning top.}. The `difficulty' gives the constant of proportionality between the length of the path as measured in the Killing metric, $s=t$, and the length of the same path as measured in the complexity geometry, $L = \Gamma s = \Gamma t$.

\subsection{Approximation 1: prune expensive operators} \label{subsec:pruningerror}

Our first approximation will be to prune the Hamiltonian of its most expensive components. We will drop all terms that have a penalty factor greater than some critical value $\overline{\mathcal{I}}$,
\begin{equation}
\textrm{approximation 1: \ \ } \ \ H(t) = \sum_I h_I(t) \sigma_I \ \rightarrow \ H_P(t) \equiv \sum_{ I \textrm{ with } \mathcal{I}(\sigma_I) < \overline{\mathcal{I}}}  h_I(t) \sigma_I \ . \label{eq:approximation1}
\end{equation} 
Excision introduces error. Let's upperbound that error. 

\subsubsection{Killing error from pruning} 
According to Eq.~\ref{eq:distancebetweentwodifferentpaths}, the Killing error from pruning is upperbounded by 
\begin{equation}
s(\mathcal{P} e^{i \int H(t) dt} ,  \mathcal{P} e^{i \int H_P(t) dt} ) \ \leq \  \int dt \sqrt{ \overline{\textrm{Tr}}[ (H(t) - H_P(t))^2  ] }   =  \int dt \sqrt{ \sum_{\mathcal{I}(\sigma_I) > \overline{\mathcal{I}}}  h_I(t)^2  }   \ .  
\end{equation}
Eqs.~\ref{eq:tracenormalization} and \ref{eq:conservedhardness} tell us that very expensive operators must also be very small, since 
\begin{equation}
\Gamma^2 = \sum_I \mathcal{I}(\sigma_I) h_I(t)^2  \geq \overline{\mathcal{I}} \sum_{\mathcal{I}(\sigma_I) > \overline{\mathcal{I}}}  h_I(t)^2  \ . 
\end{equation}
Combining these equations, the total Killing error is no more than 
\begin{equation}
s(\mathcal{P} e^{i \int H(t) dt} ,  \mathcal{P} e^{i \int H_P(t) dt} ) \ \leq \ \frac{\Gamma t}{\sqrt{\overline{ \mathcal{I}}}} =\frac{L }{\sqrt{\overline{ \mathcal{I}}}} \ . \label{eq:innerproducterrorexcision}
\end{equation}

\subsubsection{Operator-norm error from pruning} \label{subsubsec:operatornormpruning} 
In Sec.~\ref{sec:operrorpruningformula}, we will prove that the total operator-norm error from pruning is upperbounded by 
\begin{equation}
\bigl| \bigl| \mathcal{P} e^{i \int H(t) dt} -  \mathcal{P} e^{i \int H_P(t) dt} \bigl| \bigl|_{\textrm{op} }  \ \leq  \  L  \ \textrm{min} \biggl[ \sqrt{ \sum_{\mathcal{I}(\sigma_I) > \overline{\mathcal{I}}}  \frac{1}{\mathcal{I}(\sigma_I)} }  , \frac{2^{N/2}}{ \sqrt{\overline{\mathcal{I}} }} \biggl] \ . 
\end{equation}
(There are two upperbounds because there are two upperbounds in Eq.~\ref{eq:norminequalitieswithboth}.) The operator-norm error is larger than the Killing error Eq.~\ref{eq:innerproducterrorexcision}. 

\subsection{Approximation 2: average Hamiltonian} \label{subsec:approx2}

To approximate the path generated by $H_P(t)$, we will adopt the standard quantum simulation strategy  \cite{Lloyd} of divide and conquer. First we will divide the path into $S$ equal segments, each of complexity-geometry length $L/S$ and inner-product length $\delta \equiv L/\Gamma S$,
\begin{equation}
\mathcal{P}  \exp\Bigl[ i \int_0^{S \delta} \hspace{-1mm}  {H}_P(t) dt \Bigl]  =   \biggl( \mathcal{P} \exp\Bigl[ i \int^{S \delta}_{(S-1) \delta} \hspace{-1mm}   {H}_P(t) dt \Bigl]  \biggl)   \ldots  
\biggl(  \mathcal{P}  \exp\Bigl[ i \int_0^{\delta}  {H}_P(t) dt \Bigl]  \biggl)       ; 
\end{equation}
 and then we will conquer each segment in turn by approximating it within our target error. Rather than following every twist and turn of $H_P(t)$, which would be prohibitively expensive, we will instead, for each segment, just apply the  average Hamiltonian within that segment, 
\begin{equation}
\textrm{approximation 2}: \ \ \ \  \mathcal{P}  \exp\Bigl[ i \int_{T}^{T + \delta} H_P(t) dt \Bigl]  \ \  \rightarrow  \ \   \exp\Bigl[ i \int_{T}^{T + \delta} H_P(t) dt \Bigl]   \ . \label{eq:approximation2}
\end{equation}
These two unitaries agree at O($\delta$) but disagree at O($\delta^2$). Since we will eventually make $\delta$ small but non-infinitesimal, we must also be careful about terms higher order in $\delta$. We'll exercise this care in the appendix, and quote the results here. 

\subsubsection{Killing error from averaging} 
In Appendix~\ref{subsec:averagingerrorfrobeniusappendix} we will show that the per-segment Killing error caused by averaging is upperbounded by 
\begin{equation}
\delta < \mathcal{N}^{-\frac{1}{2}}  \ \ \ \ \rightarrow \ \  \ \ s(\mathcal{P}  \exp\Bigl[ i \int_{T}^{T +  \delta} H_P(t) dt \Bigl] ,  \exp\Bigl[ i \int_{T}^{ T + \delta} H_P(t) dt \Bigl]  )  \ \leq \ \pi \sqrt{\mathcal{N}} \delta^2 \ . 
 \end{equation}

\subsubsection{Operator-norm error from averaging} 
In Appendix~\ref{subsec:averagingerroroperatorappendix} we will show that the per-segment operator-norm error caused by averaging is upperbounded by
\begin{equation}
 \delta < \mathcal{N}^{-1/2} \  \ \rightarrow \ \ \ || \mathcal{P}  \exp\Bigl[ i \int_{T}^{ T + \delta} H_P(t) dt \Bigl]  -  \exp\Bigl[ i \int_{T}^{T +  \delta} H_P(t) dt \Bigl]  ||_{\textrm{op}} \ \leq \  2 \mathcal{N} \delta^2 \ .
\end{equation}

\subsection{Approximation 3: Trotterize unitary} \label{subsec:trottererror}

 The final approximation will be to chop up the average Hamiltonian into its constituent generalized Pauli's and implement each Pauli in turn. This is known as Trotterization. We will use the first-order Trotter approximation
\begin{equation}
\exp \Bigl[  i \int_{T}^{T + \delta} dt  \ \sum_{I = 1}^{\mathcal{N}} h_I(t)  {\sigma}_I  \Bigl]  \  \rightarrow \   \prod_{I = 1}^{\mathcal{N}} \exp \Bigl[ i   \int_{T}^{T + \delta}  dt \, h_I(t)  {\sigma}_I \Bigl]    \ . 
 \end{equation}
These agree at O($\delta$) but disagree at O($\delta^2$), 
\begin{equation}
 e^{ i \int_{T}^{T + \delta} dt \sum_I^{\mathcal{N}}  h_I(t)  {\sigma}_I  } -   \prod_{I=1}^\mathcal{N} \left( e^{i   \int_{T}^{T + \delta} dt \, h_I(t)  {\sigma}_I }  \right)  = \frac{1}{2}  \sum_{I= 1}^\mathcal{N} \sum_{J<I}  h_I h_J [ \sigma_I,\sigma_J] \delta^2 + O(\delta^3)  \ . 
 \end{equation}

\subsubsection{Killing error from Trotterization} 
In Appendix~\ref{appendix:trottererrorfrobeniusnorm} we will show that the per-segment Killing error caused by Trotterization is upperbounded by
\begin{equation}
s\Bigl(e^{ i \int_{T}^{T + \delta} dt \sum_I h_I(t)  {\sigma}_I  } ,  \prod_{I = 1}^\mathcal{N}   e^{i   \int_{T}^{T + \delta} dt \, h_I(t)  {\sigma}_I }  \Bigl)  \ \leq \ \frac{\pi}{2} \sqrt{\mathcal{N}} \delta^2 \ .  \label{eq:referbacktothissquareroottocastdoubtonit}
 \end{equation}

\subsubsection{Operator-norm  error from Trotterization} 
In Appendix~\ref{appendix:trottererrorperatornorm} we will show that the per-segment operator-norm error caused by Trotterization is upperbounded by
 \begin{equation}
||  e^{ i \int_{T}^{T + \delta} dt \sum_I h_I(t)  {\sigma}_I  } -   \prod_{I=1}^\mathcal{N}  e^{i   \int_{T}^{T + \delta} dt \, h_I(t)  {\sigma}_I }   ||_{\textrm{op}} \  \leq \   \delta^2 { \mathcal{N}} \ . 
 \end{equation}

\subsection{Total error from combining segments} 
Errors at-worst add for both the Killing distance (Eq.~\ref{eq:biinvariantriemanniancompose}) and the operator norm (Eq.~\ref{eq:compositionoferrorsfornorms}). The total  error is therefore no more than the sum of the error contributions for each segment times the number of segments $S = L / \Gamma \delta$.  So long as we keep $ \delta < \mathcal{N}^{-1/2}$, the results in Secs.~\ref{subsec:pruningerror}-\ref{subsec:trottererror} imply the error is upperbounded by 
\begin{eqnarray}
s(\textrm{error} ) \, & \leq &  L \left( \frac{3 \pi}{2}  \frac{\sqrt{\mathcal{N}} \delta}{\Gamma}   + \frac{1}{\sqrt{\overline{ \mathcal{I}}}}   \right) \  \label{eq:totalbiinvarianterror} \\
|| \textrm{error} ||_{\textrm{op}} & \leq & L \left( 3 \frac{\mathcal{N} \delta}{\Gamma}   + \textrm{min} \left[ \sqrt{ \sum_{\mathcal{I}(\sigma_I) > \overline{\mathcal{I}}}  \frac{1}{\mathcal{I}(\sigma_I)} }  \ , \  \frac{2^{N/2}}{\sqrt{ \overline{ \mathcal{I}}}} \right]  \right) \ .  \label{eq:totaloperatornormerrorbudget}
\end{eqnarray}
To keep the first contribution in budget, we must keep $\delta$ small; to keep the second contribution in budget, we must keep $\overline{\mathcal{I}}$ large.

\subsection{Lemma: making monomial unitaries} \label{sec:makingmonomials}
We have taken the path and broken it up into segments, and then broken those segments up into `monomial' unitaries each generated by a single generalized Pauli. Now let's show that we can make any of those monomial unitaries with a small number of gates, even if the monomial has high weight.

If $\sigma_K$ is a $k$-local generalized Pauli operator, then any `monomial' unitary of the form $e^{i \sigma_K z}$ can be synthesized exactly using no more than $2k-3$ two-local gates, 
\begin{equation}
\textrm{lemma: }\ \ \mathcal{C}_{\textrm{gates}} [e^{i \sigma_K z} ] \ \leq \ 2 \, \textrm{weight}(\sigma_K) -3 \ . \label{lemma:lemma1}
\end{equation}
Let's prove that now. Without loss of generality, we can assume the monomial has the form 
\begin{equation}
e^{i \sigma_K z} = e^{i \sigma_z \otimes \sigma_z \otimes \ldots \otimes \sigma_z \otimes \sigma_z z } \equiv e^{i (\sigma_z)_1 (\sigma_z)_2 \ldots (\sigma_z)_{k-1} (\sigma_z)_k  z }  \ .
\end{equation}
(This is without loss of generality because all monomial operators are related by a change of single-qubit basis, e.g.~changing $\sigma_z$ to $\sigma_x$ on the first qubit, changing $\sigma_z$ to $\sigma_y$ on the second qubit, etc. But this is a symmetry of our definition of gate complexity. In Sec.~\ref{subsec:gatecomplexity}, we defined the primitive gates to be \emph{any} two-qubit unitary---any element of U($4$)---so we can effect any change of single-qubit basis by just bundling it into the last gate that touches that qubit.) We will show how to write $e^{i \sigma_K z}$ with $2k-3$ gates. First, note the identity 
\begin{equation}
e^{i \sigma_z z} = e^{i \sigma_x \frac{\pi}{4}} e^{i \sigma_y z} e^{-i \sigma_x \frac{\pi}{4}} \ :
\end{equation}
to change yaw by $2z$, we can roll  $90^{\degree}$, pitch down by $2z$, and then roll back \cite{yaw}. This identity can be applied to any three generators that have the same algebra as SU(2), so for any generalized Pauli's $\sigma_I$ and $\sigma_J$ we have  
\begin{equation}
e^{i (\sigma_I \otimes \sigma_z \otimes \sigma_J) z} = e^{i (\sigma_I \otimes \sigma_x \otimes \id) \frac{\pi}{4}} e^{i (\id \otimes \sigma_y \otimes \sigma_J) z} e^{-i (\sigma_I \otimes \sigma_x \otimes \id) \frac{\pi}{4}} \ . 
\end{equation}
This allows us to recursively build up high-weight monomials 
\begin{eqnarray}
e^{i \sigma_K z} &=& e^{i (\sigma_z)_1(\sigma_x)_2 \frac{\pi}{4}} e^{i (\sigma_y)_2(\sigma_y)_3 \frac{\pi}{4}} e^{i (\sigma_x)_3(\sigma_x)_4 \frac{\pi}{4}} \ldots  e^{i (\sigma_y)_{k-3}(\sigma_y)_{k-2} \frac{\pi}{4}}    e^{i (\sigma_x)_{k-2}(\sigma_x)_{k-1} \frac{\pi}{4}} \\
&& \  e^{i (\sigma_y)_{k-1}(\sigma_z)_k z}  \nonumber \\
&& \ \    e^{-i (\sigma_x)_{k-2}(\sigma_x)_{k-1} \frac{\pi}{4}} 
e^{-i (\sigma_y)_{k-3}(\sigma_y)_{k-2} \frac{\pi}{4}}   
\ldots 
e^{-i (\sigma_x)_3(\sigma_x)_4 \frac{\pi}{4}} 
e^{-i (\sigma_y)_2(\sigma_y)_3 \frac{\pi}{4}} 
e^{-i (\sigma_z)_1(\sigma_x)_2 \frac{\pi}{4}} \nonumber
\end{eqnarray}
This $2k-3$ gate circuit exactly compiles the monomial unitary and proves Eq.~\ref{lemma:lemma1}. Since $2k-3 <2 N$, it is certainly the case that for all monomial unitaries of any weight, 
\begin{equation}
 \mathcal{C}_{\textrm{gates}} [e^{i \sigma_K z} ] \ < \ 2N \ . 
\end{equation}

 \subsection{Main results: bounding the gate complexity} \label{subsec:mainresults}
The number of gates to synthesize a single segment using the strategy in Sec.~\ref{sec:makingmonomials} is 
 \begin{equation}
\mathcal{C}_\textrm{gates} \left[ \prod_{I \textrm{ with } \mathcal{I}(\sigma_I) < \overline{\mathcal{I}}}   \left( e^{i   \int_{T}^{T + \delta} dt \, h_I(t)  {\sigma}_I }  \right) \right] < 2 N  \mathcal{N} \ . 
 \end{equation}
The total cost is the cost-per-segment times the number of segments $S = \frac{L}{\Gamma  \delta}$, 
 \begin{equation}
\mathcal{C}_\textrm{gates} \left[  \prod_\textrm{segments} \prod_{I \textrm{ with } \mathcal{I}(\sigma_I) < \overline{\mathcal{I}}}  \left( e^{i   \int_{T}^{T + \delta} dt \, h_I(t)  {\sigma}_I }  \right) \right] \  <   \ 2 N  S  \,  \mathcal{N}  =  \frac{2 N  L  \,  \mathcal{N}}{\Gamma \delta} \ . \label{eq:totalgatecostformula}
 \end{equation}

 \subsubsection{The gate complexity of getting Killing close} 
Our error budget is set by Eq.~\ref{eq:totalbiinvarianterror}. To stay within this budget, we will ensure that each of the two terms in the sum in Eq.~\ref{eq:totalbiinvarianterror} is less than half the permitted error; apportioning half the budget to each term and then plugging into Eq.~\ref{eq:totalgatecostformula}  gives 
\begin{equation}
\delta < \frac{\Gamma}{3 \pi  \sqrt{\mathcal{N}} L }  s(\textrm{error}) \ \ \& \ \  \frac{1}{ \sqrt{\overline{ \mathcal{I}}}}   < \frac{s(\textrm{error})}{2 L   }  \ \ \rightarrow \ \ \mathcal{C} < \frac{6 \pi N \mathcal{N}^\frac{3}{2}  }{\Gamma^2 }  \frac{L^2}{ s(\textrm{error}) } \ . 
\end{equation} 
This first inequality (together with the fact that unless $L > \Gamma s(\textrm{error})$ the two unitaries to be connected are already within $s(\textrm{error})$ at the start) ensures that $\delta < \mathcal{N}^{- 1/2}$ and therefore self-consistently ensures the validity of Eq.~\ref{eq:totalbiinvarianterror}. Since $\Gamma^2 \geq \textrm{min}_{I} \mathcal{I}(\sigma_I) = 1$ for every geodesic, this tells us that whenever there is a complexity geometry path of length $L$, there is also a circuit that arrives within Killing distance $s(\textrm{error})$ that uses a number of two-local gates no more than 
 \begin{equation}
\boxed{ \mathcal{C}_\textrm{gates} <  \frac{ 6 \pi N \mathcal{N}_{\overline{\mathcal{I}}}^{3/2}  L^2}{ s(\textrm{error}) }  \  \ \ \textrm{ where } \mathcal{N}_{\overline{\mathcal{I}}} \textrm{ is the number of $\sigma_I$ with } \mathcal{I}(\sigma_I) \leq \overline{\mathcal{I}} = \frac{4 L^2 }{  s(\textrm{error})^2  } } \  . \label{eq:boxedgettingFrobeniusclose}
\end{equation}
A consequence of this inequality is the criterion for polynomial equivalence, Eq.~\ref{sufficientcondition1}.

 \subsubsection{The gate complexity of getting operator-norm close} 
Our error budget is set by Eq.~\ref{eq:totaloperatornormerrorbudget}.  Let's first use the bound from the first term in the minimum in Eq.~\ref{eq:totaloperatornormerrorbudget}. Apportioning half the budget to each of the two terms in the sum and then plugging into Eq.~\ref{eq:totalgatecostformula}  gives 
\begin{equation}
\delta < \frac{\Gamma}{6 {\mathcal{N}} L }  || \textrm{error} ||_\textrm{op} \ \ \& \ \  \sum_{\mathcal{I}(\sigma_I) > \overline{\mathcal{I}}}  \frac{1}{\mathcal{I}(\sigma_I)}   < \frac{|| \textrm{error} ||_\textrm{op}^2}{4 L^2   }  \ \ \rightarrow \ \ \mathcal{C} < \frac{12N \mathcal{N}_{\overline{\mathcal{I}}}^2  }{\Gamma^2 }  \frac{L^2}{ || \textrm{error} ||_\textrm{op}} \ .  \label{eq:errorapportionmentoperatornorm}
\end{equation} 
Using $\Gamma \geq 1$, and recalling that $\mathcal{N}_{\overline{\mathcal{I}}}$ is defined as the number of $\sigma_I$ with $\mathcal{I}(\sigma_I) \leq \overline{\mathcal{I}}$, 
\begin{equation}
\boxed{ \mathcal{C}_\textrm{gates} <  \frac{ 12 N \, \mathcal{N}_{\overline{\mathcal{I}}}^{\, 2}  \,  L^2}{ || \textrm{error} ||_\textrm{op}}   \textrm{ where } \overline{\mathcal{I}} \textrm{ is big enough that } \hspace{-4mm}  \sum_{I \textrm{ with } \mathcal{I}(\sigma_I) > \overline{\mathcal{I}}}  \frac{1}{\mathcal{I}(\sigma_I)}   \leq \frac{  || \textrm{error} ||_\textrm{op}^2  }{4 L^2 } }   \  .  \label{eq:boxedgettingoperatorclose}
\end{equation}
A consequence of this inequality is the criterion for polynomial equivalence, Eq.~\ref{sufficientcondition2}. 
We can also derive the bound from the second term in the minimum in Eq.~\ref{eq:totaloperatornormerrorbudget}. This gives the sometimes-tighter-sometimes-looser 
\begin{equation}
\boxed{\mathcal{C}_\textrm{gates} <  \frac{ 12 N \, \mathcal{N}_{\overline{\mathcal{I}}}^{\, 2}  \,  L^2}{ || \textrm{error} ||_\textrm{op}} \  \ \ \textrm{ where } \mathcal{N}_{\overline{\mathcal{I}}} \textrm{ is the number of $\sigma_I$ with } \mathcal{I}(\sigma_I) \leq \overline{\mathcal{I}} = 2^N\frac{ 4 L^2 }{  ||\textrm{error}||_\textrm{op}^2  }  } \ . \label{eq:boxedgettingoperatorclose2}
\end{equation} 
A consequence of this inequality is the criterion for polynomial equivalence, Eq.~\ref{sufficientcondition3}. \\

\noindent Comparing Eq.~\ref{eq:boxedgettingFrobeniusclose} with Eqs.~\ref{eq:boxedgettingoperatorclose}-\ref{eq:boxedgettingoperatorclose2} we see that it is easier to get Killing close, reflecting the fact that the Killing distance is generally shorter than the operator-norm distance.

\section{Bounds for example metrics} 
Let's illustrate our results by taking the general bounds Eqs.~\ref{eq:boxedgettingFrobeniusclose}, \ref{eq:boxedgettingoperatorclose}, and \ref{eq:boxedgettingoperatorclose2} and applying them to the example metrics Eqs.~\ref{eq:examplemetric1}-\ref{eq:examplemetric3}. We will also have the trivial bound from Eq.~\ref{eq:trivialbound}, 
\begin{equation}
\mathcal{C}_\textrm{gates} \leq N^2 4^N \ . \label{eq:trivialbound2}
\end{equation}

\subsection{Bounds for Killing close} 
The number of gates required to get Killing close is bounded by  Eq.~\ref{eq:boxedgettingFrobeniusclose}. (As discussed around Eq.~\ref{eq:tobeprovedinappendix1}, being Killing close is equivalent to being close in the normalized Frobenius-norm distance.)   

\subsubsection{Cliff metric, Killing close}
The cliff metric is 
 \begin{equation}\mathcal{I}_{k\leq2} = 1, \ \mathcal{I}_{k\geq3} = \mathcal{I}_\textrm{cliff} \ . 
\end{equation}
For the cliff metric, the number of cheap direction is $\mathcal{N}_2 + \mathcal{N}_1 + \mathcal{N}_0$ so Eq.~\ref{eq:boxedgettingFrobeniusclose} gives 
\begin{equation}
\frac{4 L^2 }{  s(\textrm{error})^2  }  \leq {\mathcal{I}_\textrm{cliff}} \ \ \ \ \rightarrow \ \ \  \ \mathcal{C}_\textrm{gates} <  6 \pi N \left( \frac{9 N^2 - 3N + 2}{2} \right) ^\frac{3}{2}   \frac{   L^2}{ s(\textrm{error})} \ . \label{eq:boundforfrobeniuscliff}
\end{equation}
The combination of Eqs.~\ref{eq:trivialbound2} and \ref{eq:boundforfrobeniuscliff} tells us that so long as $\mathcal{I}_\textrm{cliff}$ is exponentially large\footnote{At the risk of being obtuse, one could also consider values of $\mathcal{I}_\textrm{cliff}$ that are super-polynomial but sub-exponential in $N$, in which case Eq.~\ref{eq:boundforfrobeniuscliff} holds for all \emph{polynomial} $L$s, but there is not a polynomial bound on $\mathcal{C}_\textrm{gates}$ for all $L$ and $s(\textrm{error})$.} (meaning that $\mathcal{I}_\textrm{cliff} > \alpha^N$ for some $\alpha > 1$) then 
\begin{eqnarray}
\textrm{exponential $\mathcal{I}_\textrm{cliff}$} &\rightarrow &  \textrm{Eq.~\ref{eq:boundforfrobeniuscliff} holds for all \emph{polynomial} values of $L$ and $s(\textrm{error})$}  \\
& \rightarrow &  \textrm{$\mathcal{C}_\textrm{gates}$ is a polynomial function of $L$ and $s(\textrm{error})$ for \emph{all}}   \\
&& \textrm{values of $L$ and $s(\textrm{error})$} \nonumber 
\end{eqnarray}
For $\mathcal{I}_\textrm{cliff} > 4^N$,   Eq.~\ref{eq:boundforfrobeniuscliff} holds for all $L$ and $s(\textrm{error})$.

\subsubsection{Binomial metric, Killing close}
For the binomial metric the penalty factor of a $k$-local direction is (the $\alpha$th power of) the number of $k$-local directions, 
 \begin{equation}
 \mathcal{I}_{k} = (\mathcal{N}_k)^\alpha \equiv \left( {N \choose k} 3^k \right)^\alpha \ . 
\end{equation}
Our bound Eq.~\ref{eq:boxedgettingFrobeniusclose}  gives 
\begin{equation}
 \mathcal{C}_\textrm{gates} <  \frac{ 6 \pi N \mathcal{N}_k^\frac{3}{2}  L^2}{ s(\textrm{error}) }  \  \ \ \textrm{ where $k$ is big enough that } (\mathcal{N}_k)^\alpha \geq \frac{4 L^2 }{ s(\textrm{error})^2  } \  .
\end{equation}
This tells us that  
\begin{equation}
\mathcal{C}_\textrm{gates} <  \frac{ 6 \pi N   L^2}{  s(\textrm{error})}  \left(  \frac{2L}{s(\textrm{error})}  \right)^{\frac{3}{ \alpha}} \ . 
\end{equation}
So long as $\alpha>0$, this gives polynomial equivalence for any $L$ and $s(\textrm{error})$.

\subsubsection{Exponential metric, Killing close}
For the exponential metric, the penalty factor is exponential in the $k$-locality, 
 \begin{equation}
 \mathcal{I}_{k} = x^{2k} \ .
\end{equation}
Our bound Eq.~\ref{eq:boxedgettingFrobeniusclose}  gives 
\begin{equation}
 \mathcal{C}_\textrm{gates} <  \frac{ 6 \pi N \mathcal{N}_k^\frac{3}{2}  L^2}{s(\textrm{error}) }  \  \ \ \textrm{ where $k$ is big enough that } x^{2k}  \geq \frac{4 L^2  }{  s(\textrm{error})^2  } \  .
\end{equation}
This is not \emph{polynomial }equivalence, since the bound on the number of gates scales like $L^{\log N/\log x}$, and so as $N$ increases the exponent of $L$ grows without bound. However, because $\log N$ grows so slowly, for all $x>1$ this gives what is known as 
 \emph{quasi}-polynomial equivalence.

\subsection{Bounds for operator-norm close} 
The number of gates required to get operator-norm close is bounded by  Eqs.~\ref{eq:boxedgettingoperatorclose} and \ref{eq:boxedgettingoperatorclose2}. 
\subsubsection{Cliff metric, operator-norm close}
The cliff metric is 
 \begin{equation}\mathcal{I}_{k\leq2} = 1, \ \mathcal{I}_{k\geq3} = \mathcal{I}_\textrm{cliff} \ . 
\end{equation}
For the cliff metric, the strongest bound comes from Eq.~\ref{eq:boxedgettingoperatorclose2}, which gives 
\begin{equation}
\frac{4 L^2 }{  || \textrm{error} ||_{ \textrm{op}}^2  }  \leq 2^{-N} \mathcal{I}_\textrm{cliff}   \ \ \ \rightarrow \ \ \    \ \mathcal{C}_\textrm{gates} <  12 N \left( \frac{9 N^2 - 3N + 2}{2} \right) ^{2}   \frac{   L^2}{ || \textrm{error} ||_{\textrm{op}}} \ . \label{eq:answerforcliffoperatorclose}
\end{equation}
The combination of Eqs.~\ref{eq:trivialbound2} and \ref{eq:answerforcliffoperatorclose} tells us that so long as $2^{-N}{\mathcal{I}_\textrm{cliff}}{}$ is exponentially large (meaning that $2^{-N} \mathcal{I}_\textrm{cliff} > \alpha^N$ for some $\alpha > 1$) then 
\begin{eqnarray}
\textrm{exponential $2^{-N} {\mathcal{I}_\textrm{cliff}}$} &\rightarrow &  \textrm{Eq.~\ref{eq:answerforcliffoperatorclose} holds for all \emph{polynomial} values of $L$ and $|| \textrm{error}||_{\textrm{op}}$} \nonumber  \\
& \rightarrow &  \textrm{$\mathcal{C}_\textrm{gates}$ is a polynomial function of $L$ and $|| \textrm{error}||_{\textrm{op}}$ for \emph{all}} \nonumber  \\
&& \textrm{values of $L$ and $|| \textrm{error}||_{\textrm{op}}$}  
\end{eqnarray}
For $2^{-N} \mathcal{I}_\textrm{cliff} > 4^N$,   Eq.~\ref{eq:answerforcliffoperatorclose} holds for all $L$ and $|| \textrm{error}||_{\textrm{op}}$.

\subsubsection{Binomial metric, operator-norm close} \label{subsubsec:binomialmetricoperatorclose}
For the binomial metric the penalty factor of a $k$-local direction is (the $\alpha$th power of) the number of $k$-local directions, 
 \begin{equation}
 \mathcal{I}_{k} = (\mathcal{N}_k)^\alpha \equiv \left( {N \choose k} 3^k \right)^\alpha \ . 
\end{equation}
For the binomial metric, the strongest bound comes from Eq.~\ref{eq:boxedgettingoperatorclose}. 
Eq.~\ref{eq:errorapportionmentoperatornorm} tells us that to control the pruning error we need 
\begin{equation}
 \sum_{I \textrm{ with } \mathcal{I}(\sigma_I) > \overline{\mathcal{I}}}  \frac{1}{\mathcal{I}(\sigma_I)}    =  \sum_{k \textrm{ with } \mathcal{N}_k > \mathcal{N}_{\bar{k}}}    \mathcal{N}_k^{1-\alpha}   \leq \frac{  || \textrm{error} ||_\textrm{op}^2  }{4 L^2 }  \ . 
 \end{equation} 
For $\alpha \geq 1$ we have 
\begin{equation}
 \sum_{k \textrm{ with } \mathcal{N}_k > \mathcal{N}_{\bar{k}}}    \mathcal{N}_k^{1-\alpha}  \leq   N \mathcal{N}_{\bar k}^{1-\alpha}  \ . 
 \end{equation} 
Thus for $\alpha > 1$,  Eq.~\ref{eq:boxedgettingoperatorclose} tells us that polynomial $L$ and $|| \textrm{error} ||_\textrm{op}$ means polynomial $\mathcal{C}_\textrm{gates}$, 
\begin{equation}
 \mathcal{C}_\textrm{gates} <  \frac{ 12 N   L^2}{ || \textrm{error} ||_{ \textrm{op}}} \left(  \frac{4 N L^2 } {  || \textrm{error} ||_\textrm{op}^2  }\right)^{\frac{2}{\alpha - 1} }  \  \ . \label{eq:alphaisoneisspecial}
 \end{equation}
We have no polynomial bound for $\alpha \leq 1$.
 
\subsubsection{Exponential metric, operator-norm close}
For the exponential metric, the penalty factor is exponential in the $k$-locality, 
 \begin{equation}
 \mathcal{I}_{k} = x^{2k} \ .
\end{equation}
Eq.~\ref{eq:errorapportionmentoperatornorm} tells us that to control the pruning error we need 
\begin{equation}
 \sum_{I \textrm{ with } \mathcal{I}(\sigma_I) > \overline{\mathcal{I}}}  \frac{1}{\mathcal{I}(\sigma_I)}  = \sum_{k > \bar{k}}  \mathcal{N}_k x^{-2k}    \leq \frac{  || \textrm{error} ||_\textrm{op}^2  }{4 L^2 }  \ . 
 \end{equation} 
 The sum is exponentially large for $x = $ O(1) unless $\mathcal{N}_{\bar{k}}$ is exponentially large, so our method does not put tight bounds on the gate complexity of following polynomial-length paths in the exponential metric if we insist on getting operator-norm close. 

\section{State complexity} \label{sec:statecomplexity}
So far in this paper we have been discussing the complexity of unitary \emph{operators}. One can also define the relative complexity of  pairs of quantum \emph{states}. The relative complexity of a pair of states is defined as the complexity of the least complex unitary that connects them, 
\begin{equation}
\mathcal{C}_\textrm{state}[ |\psi_1 \rangle ; |\psi_0 \rangle  ] \equiv \textrm{min}_U \mathcal{C}_\textrm{unitary}[U,\mathds{1}] \ \ \textrm{where} \ \  |\psi_1 \rangle = U |\psi_0 \rangle. \label{eq:statecomplexitydefinition}
\end{equation}
Analogously one can define a state-space complexity geometry (see e.g.~Sec.~3 of \cite{Brown:2019whu}). Starting with a complexity geometry on unitaries, the  distance between a pair of states is defined as the distance from the origin of the least distant unitary that connects them
\begin{equation}
{L}_\textrm{state}[ |\psi_1 \rangle ; |\psi_0 \rangle  ] \equiv \textrm{min}_U {L}_\textrm{unitary}[U,\mathds{1}] \ \ \textrm{where} \ \  |\psi_1 \rangle = U |\psi_0 \rangle. \label{eq:statecomplexitymetricdefinition}
\end{equation}
A penalty matrix $\mathcal{I}_{IJ}$ thus defines a deformed metric on Hilbert space.

It follows from   Eq.~\ref{eq:clifflessthanpigates} that the relative gate complexity of a pair of states is \emph{lower}bounded by their separation in the state-space complexity geometry of the cliff metric, or of any metric less punitive than the cliff metric, 
\begin{equation}
    \mathcal{C}_\textrm{gates}[| \psi_1 \rangle,| \psi_0 \rangle] \ \geq \ \pi^{-1} L_\textrm{cliff}[| \psi_1 \rangle,| \psi_0 \rangle] \ .  
\end{equation}
To develop a useful \emph{upper}bound, we're going to have to tolerate error. 

If we are aiming for $|\psi_a \rangle = U_a |\psi_0 \rangle$ but we instead hit $|\psi_b \rangle = U_b |\psi_0 \rangle$, then knowing that $||U_a -U_b||_{\overline{F}}$ is small (or equivalently that $U_a$ and $U_b$ are close in the Killing metric) does not guarantee that the inner-product $\langle \psi_a|\psi_b \rangle$ will be close to one. This is because small $||U_a - U_b||_{\overline{F}}$ only guarantees that $\langle \psi | U_a^\dagger U_b|\psi \rangle$ is close to one when averaged over all input states $|\psi \rangle$, and not necessarily for our particular input state $|\psi_0\rangle$. Instead  we need to demand that $U_a$ and $U_b$ are close in the operator norm. The operator norm being small does guarantee that $U_a|\psi \rangle$ and $U_b|\psi \rangle$ are close even for the worst-case input. Using the definition of the operator norm, Eq.~\ref{eq:operatornormdefinition}, we have 
\begin{equation}
\epsilon \ \equiv \   \sqrt{ 2 - 2  \textrm{Re}[\langle \psi_a | \psi_b \rangle] } \ \leq \  ||U_a - U_b||_\textrm{op} \ . \label{eq:innerproductintermsofgatecomplexity}
\end{equation}
Using this inequality, bounds on the number of gates required to get operator-norm close to a unitary become bounds on the number of gates required to get inner-product close to a state.

The first upperbound on the cost of getting operator-norm close was  Eq.~\ref{eq:boxedgettingoperatorclose}. Combining this with  Eq.~\ref{eq:innerproductintermsofgatecomplexity} upperbounds the size of the circuit need to transition between any pair of states that are connected by a complexity-geometry path of length $L$: there must exist a circuit of size at most 
\begin{equation}
\mathcal{C}_\textrm{gates} [|\psi_1 \rangle,|\psi_2\rangle ] \ < \   \frac{ 12 N \, \mathcal{N}_{\overline{\mathcal{I}}}^{\, 2}  \   L[|\psi_1 \rangle,|\psi_2\rangle]^2}{  \epsilon }   \textrm{ where } \overline{\mathcal{I}} \textrm{ is big enough that }  \sum_{\mathcal{I}(\sigma_I) > \overline{\mathcal{I}}}  \frac{1}{\mathcal{I}(\sigma_I)}   \leq \frac{  \epsilon^2  }{4 L^2 }        \label{eq:statespacebound1}
\end{equation} 
that, when applied to the first state, gets within inner-product error $\epsilon$ of the second state. 
(Recall that $\mathcal{N}_{\overline{\mathcal{I}}}$ is defined as the number of principal directions $\sigma_I$ with $\mathcal{I}(\sigma_I) \leq \overline{\mathcal{I}}$.)  
The second upperbound comes from Eq.~\ref{eq:boxedgettingoperatorclose2} and gives 
\begin{equation}
\mathcal{C}_\textrm{gates} [|\psi_1 \rangle,|\psi_2\rangle ] \ < \   \frac{ 12 N \, \mathcal{N}_{\overline{\mathcal{I}}}^{\, 2}  \   L[|\psi_1 \rangle,|\psi_2\rangle]^2}{ \epsilon }  \textrm{ where } \overline{\mathcal{I}} \  = \    2^N  \frac{ 4 L^2 }{  { \epsilon^2 }  } \ . \label{eq:statespacebound2}
\end{equation} 

All penalty schedules that are sufficiently punitive to satisfy either of the sufficient conditions Eqs.~\ref{sufficientcondition2} or \ref{sufficientcondition3} are sufficiently punitive that any pair of states that can be connected by a state-space complexity-geometry path of at most polynomial length $L[|\psi_1 \rangle,|\psi_2\rangle]$ can also be approximated to within any polynomially small inner-product error $\epsilon$ by a circuit with at most polynomially many gates.

\section{Implications for diameter of complexity geometry} 
The ``diameter'' of a space is defined as the greatest separation between any pair of points. Let's use our results to lowerbound the diameter of the complexity geometry. \\

\noindent One question that has arisen repeatedly in the literature is which complexity geometries  have a diameter that is exponential in the number of qubits $N$. It follows implicitly from the work of Nielsen \emph{et al.}~\cite{Nielsen2} that both the cliff metric for the unitary group and the cliff metric for state space have exponential diameters. Then in Ref.~\cite{Brown:2021euk} I proved that a much broader class of metrics on the unitary group have exponential diameters. In the analysis below I will show that Eq.~\ref{eq:boxedgettingFrobeniusclose} also implies an exponential diameter for a broad class of complexity geometries on the unitary group (though as we will see a slightly smaller class than was proved  to have exponential diameter in \cite{Brown:2021euk}); I will also show that Eq.~\ref{eq:statespacebound1} can be used to establish---for the first time in the literature---an exponential diameter for a broad class of complexity geometries on state space.\\

\noindent In the rest of this paper, all equalities and inequalities are exact. In this section we will only ask about the part of the answer that is exponential in $N$, and so will ignore all factors sub-exponential in $N$. For this reason, we will write `$\sim$' and `$\lsim$' not `$=$' and `$<$'.

\subsection{Diameter of complexity geometry of unitary group} \label{subsec:implicationsfordiameter}
 The main results of this paper---given in Sec.~\ref{subsec:mainresults}---are inequalities that upperbound the number of gates $\mathcal{C}_\textrm{gates}$ needed to approximate any unitary that can be reached by a path of length $L$. Alternatively, if we already know $\mathcal{C}_\textrm{gates}$, the inequalities lowerbound $L$.  Using this, let's lowerbound the  $L$ needed to approximate a \emph{typical}  unitary.

The unitary group on $N$ qubits, U($2^N$), is $4^N$-dimensional and a simple counting argument \cite{Knill:1995kz}  tells us the number of gates needed to get close to a {typical} element is  
\begin{equation}
\mathcal{C}_\textrm{gates}[U_\textrm{typical},U_\textrm{typical'}] \ \sim \  4^N \ . \label{eq:Ctypical}
\end{equation}
(Here and in the rest of this section we write ``$\sim$'' not ``$=$'' because we have neglected all sub-exponential factors.)  
 This is true  
  no matter whether we measure `closeness' in the Killing metric or in the operator norm, so to get the tightest bound on $L_\textrm{typical}$ we should use the Killing bound. Putting e.g.~$s($error$) = 10^{-2}$ into Eq.~\ref{eq:boxedgettingFrobeniusclose} and dropping all subexponential terms gives 
 \begin{equation}
 \mathcal{C}_\textrm{gates} \ \lsim \  \mathcal{N}_{\overline{\mathcal{I}}}^{3/2}  L^2   \  \ \ \textrm{ where } \mathcal{N}_{\overline{\mathcal{I}}} \textrm{ is the number of $\sigma_I$ with } \mathcal{I}(\sigma_I) \leq \overline{\mathcal{I}} \  \lsim \  L^2 \  .  
\end{equation}
 Combining this with Eqs.~\ref{eq:lessthanpiImax} and~\ref{eq:Ctypical} gives,  for each value of $\overline{\mathcal{I}}$,  a lowerbound on the length of the complexity path required to reach a typical operator
\begin{equation}
L_\textrm{typical} \ \gsim \  \sqrt{ \textrm{min} [ 4^N\mathcal{N}_{\overline{\mathcal{I}}}^{-3/2},\overline{\mathcal{I}} \, ] }  \ . 
\end{equation}
To get the tightest bound we can maximize over the choice of $\overline{\mathcal{I}}$, 
\begin{equation}
L_\textrm{typical} \ \gsim \ \textrm{max}_{\overline{\mathcal{I}}} \sqrt{ \textrm{min} [ 4^N\mathcal{N}_{\overline{\mathcal{I}}}^{-3/2},\overline{\mathcal{I}} \, ] }  \ .  \label{eq:typicalLbound}
\end{equation}
The greatest separation between a pair of unitaries in the complexity geometry---the `diameter'---must be at least as big as the typical separation, so this result also bounds the diameter. Applying this to the three example metrics Eqs.~\ref{eq:examplemetric1}-\ref{eq:examplemetric3} gives 
\begin{eqnarray}
\textrm{cliff metric } & \rightarrow &  L_\textrm{typical} \ \gsim \ \textrm{min}[2^N, \mathcal{I}_\textrm{cliff}^\frac{1}{2}] \label{eq:diameterofcliff} \\  
\mathcal{I}_k = (\mathcal{N}_k)^\alpha &  \rightarrow &  L_\textrm{typical}  \  \gsim \ \left( 2^N \right)^{ \frac{2 \alpha}{3 + 2 \alpha}}  \\
\mathcal{I}_k = x^{2k}  &  \rightarrow &  L_\textrm{typical} \ \gsim \ \ x^{\bar{k}} \ \textrm{ where }  \mathcal{N}_{\bar{k}}^{3/2} x^{2\bar{k}} \geq 4^N  \ .
\end{eqnarray}
For all three, the diameter is exponential in the number of qubits $N$ (assuming exponential $\mathcal{I}_\textrm{cliff}$, $\alpha>0$, and $x>1$).

The bound on $L_\textrm{typical}$, Eq.~\ref{eq:typicalLbound}, is able to establish an exponential diameter even for some complexity metrics for which we were unable to prove polynomial equivalence in Sec.~\ref{sec:sec3}. For example, Eq.~\ref{eq:typicalLbound} establishes that the `delayed cliff' metric defined by $\mathcal{I}_{k <  N/4} = 1$ \& $\mathcal{I}_{k \geq  N/4} = 4^N$ has exponential diameter, even though for this metric 
$\mathcal{C}_\textrm{gates}$ is not always a polynomial function of $L$ and $s(\textrm{error})$. This is because to prove that the  diameter is exponentially large we do not need to establish polynomial equivalence for all unitaries, only for typical unitaries.

In Ref.~\cite{Brown:2021euk} I lowerbounded the diameter of the complexity geometry using a completely different approach that made no reference to gates, error budgets, or $\epsilon$-balls. This bound was almost identical to Eq.~\ref{eq:typicalLbound}, except with the ${\mathcal{N}_{\overline{\mathcal{I}}}}^{-3/2}$ tightened to ${\mathcal{N}_{\overline{\mathcal{I}}}}^{-1}$. In Sec.~\ref{subsec:prospectsforimproving}, I will speculate on how the method in this paper might be improved to also yield ${\mathcal{N}_{\overline{\mathcal{I}}}}^{-1}$. It is intriguing that these two dramatically different methods give almost the same bound.

\subsection{Diameter of complexity geometry of Hilbert space}

Because of the minimization called for in Eq.~\ref{eq:statecomplexitymetricdefinition}, it does not follow that just because the complexity geometry of the unitary group has a large diameter, the corresponding Hilbert-space complexity geometry must too. As a close analogy, consider the relative computational complexity of pairs of \emph{classical} bit strings. To implement a generic Boolean function needs exponentially many gates, but any specific pair of bit strings can be connected with at most $N$ NOT gates. Similarly, we will see that there are some penalty schedules for which we can prove the unitary group has exponential diameter but for which we cannot prove the state-space complexity geometry has exponential diameter.

The Hilbert space of $N$ qubits is $2^N$-dimensional and a simple counting argument tells us that the number of gates needed to get from one typical state to inner-product close to another is    
\begin{equation}
\textrm{typical: }  
\ \ \ \ \mathcal{C}_\textrm{gates} [ |\psi_{\textrm{typical}} \rangle, |\psi_{\textrm{typical'}} \rangle] \ \sim \  2^N \ .  \label{eq:Ctypicalapprox}
\end{equation} 
(We write ``$\sim$'' not ``$=$'' because we have again neglected all sub-exponential factors.) Our first bound on the diameter of Hilbert space comes from Eq.~\ref{eq:statespacebound1}. Putting e.g.~$\epsilon = 10^{-2}$ into Eq.~\ref{eq:statespacebound1}, combining with Eq.~\ref{eq:Ctypicalapprox}, and then dropping all subexponential terms gives 
\begin{equation}
\textrm{diameter} \ \gsim \ \textrm{max}_{\overline{\mathcal{I}}}  \textrm{min} \Biggl[\, \frac{2^\frac{N}{2}}{\mathcal{N}_{\overline{\mathcal{I}}}}\, , \   \biggl( \hspace{1pt}   \sum_{\mathcal{I}(\sigma_I) > \overline{\mathcal{I}}}  \frac{1}{\mathcal{I}(\sigma_I)}  \biggl)^{-\frac{1}{2}}   \, \Biggl]  \ .  \label{eq:diameterstatespacefirstbound}
\end{equation}
The second bound comes from Eq.~\ref{eq:statespacebound2}.  Putting e.g.~$\epsilon = 10^{-2}$ into Eq.~\ref{eq:statespacebound2}, combining with Eq.~\ref{eq:Ctypicalapprox} and then dropping all subexponential terms gives 
\begin{equation}
\textrm{diameter} \ \gsim \ \textrm{max}_{\overline{\mathcal{I}}}  \textrm{min} \left[ \, \frac{2^\frac{N}{2} }{\mathcal{N}_{\overline{\mathcal{I}}}} \, , \,  \left( \frac{ {\overline{\mathcal{I}} }}{2^{N}} \right)^\frac{1}{2} \, \right]  \ . 
\end{equation}
Applying these bounds to the cliff metric Eq.~\ref{eq:examplemetric1} and the binomial metric Eq.~\ref{eq:examplemetric2} gives 
\begin{eqnarray}
\textrm{cliff metric } & \rightarrow &  L_\textrm{typical} \ \gsim \ \textrm{min}[2^{\frac{N}{2}}, {2^{-\frac{N}{2}}} {\mathcal{I}_\textrm{cliff}^\frac{1}{2}} ] \label{eq:diameterofcliffstate} \\  
\ \ \ \ \ \ \ \ \ \ \ \ \ \ \ \ \ \ \mathcal{I}_k = (\mathcal{N}_k)^\alpha &  \rightarrow &  L_\textrm{typical}  \  \gsim \ \left( 2^\frac{N}{2} \right)^{ \frac{\alpha - 1}{  \alpha + 1 }}   \ \ \  \ \ \ \ \ \ \  (\textrm{for } \alpha \geq 1)
\end{eqnarray}
For both of these metrics, the sufficient conditions for exponential diameter (exponentially large $2^{-N} \mathcal{I}_\textrm{cliff}$, and $\alpha>1$ respectively) are the same as the sufficient conditions for polynomial equivalence of polynomial lengths, Eqs.~\ref{sufficientcondition2} and~\ref{sufficientcondition3}. By contrast for no O(1) value of $x$ does the exponential metric meet sufficient conditions Eqs.~\ref{sufficientcondition2} or \ref{sufficientcondition3}, but nevertheless for $x > 10.27...$ our bounds show that the diameter of the state-space complexity geometry is exponentially large. For $x > 10.27...$, Eq.~\ref{eq:diameterstatespacefirstbound} lowerbounds the diameter by 
\begin{eqnarray}
\ \ \ \ \ \ \mathcal{I}_k = x^{2{k}}  &  \rightarrow &  L_\textrm{typical} \ \gsim \ \ 2^{-N/2} x^{\bar{2k}} \ \textrm{ where }  \mathcal{N}_{\bar{k}} x^{2\bar{k}} = 2^N  \ .
\end{eqnarray} \\

\noindent We have just proved a theorem about the diameter of the  complexity geometry. Even though the theorem itself knows only about the complexity metric, the proof leans heavily on the Killing metric. It is curious that the proof needed to make use of this crutch, and curious that this crutch is so physically motivated. Had we just been presented with the complexity metric, with no mention of the Killing metric, how would we have known that inventing a fictitious metric would prove so useful?

\section{Discussion}

In this paper we have proved that a broad class of complexity geometries have a computational power that  is polynomially equivalent to  quantum circuits. A summary of our results is given in Sec.~\ref{subsec:summaryofresults}.

\subsection{Overview of method}
 Let's review how we proved these equivalencies. 

One direction of the polynomial equivalence is straightforward: it is easy to show that the complexity geometries are no \emph{less} powerful than circuits. This is because any quantum circuit composed of two-local gates describes a zig-zagging path through the unitary group that moves only in one- and two-local directions. Since moving in one- and two-local directions is cheap for all the complexity geometries in our universality class, it is cheap to follow the zig-zag path exactly (see Eq.~\ref{eq:clifflessthanpigates}). Error need not be tolerated. 

The nontrivial direction---and the principal result of this paper---is proving that the complexity geometries are no \emph{more} powerful than quantum gates. The problem is that if we want to use gates to \emph{exactly} follow every twist and turn of a geodesic through the complexity geometry, it is prohibitively and usually infinitely expensive. Let's describe the aspects of this problem, and how they are solved by our approximations. 

 A geodesic through the complexity geometry is generated by a Hamiltonian $H(t) = \sum_I h_I(t) \sigma_I$. By contrast for circuits the primitive operation is acting with a gate, which can enact any element of U(4) on any pair of qubits. Each gate costs 1, no matter which element of U(4) we use. As we saw in Sec.~\ref{sec:makingmonomials}, by composing these gates we can implement any monomial unitary $e^{i \sigma_I \delta}$, for any generalized Pauli $\sigma_I$ and any $\delta$,  at modest cost. 
 
 The most severe problem with trying to use gates to exactly follow a geodesic is the `microstep' problem. In the complexity geometry, the cost to implement a small step $e^{i \sigma_I \delta}$ scales like $\delta$, and so is cheap for small $\delta$. Geodesics can change direction with no penalty. Conversely the number of gates required to implement $e^{i \sigma_I \delta}$ is never less than one, no matter how small the nonzero value of $\delta$.  There is no discount for small $\delta$. This means that to exactly follow a geodesic described by a time-dependent Hamiltonian requires an infinite number of gates. This forces us to abandon attempting to follow the geodesic exactly, and instead settle for approximating the geodesic as best we can with zig-zag segments. These segments have to be short for the approximation to be accurate, but shorter segments mean more gates. 

But the time-dependence of $H(t)$ is not the only problem. Even when the Hamiltonian that generates the geodesic is independent of time, gates still struggle to keep up. This is due to the `Pythagoras' problem. When gates implement a monomial $e^{i \sigma_I \delta}$, they move in a cardinal direction in the tangent space. But when geodesics move through the unitary group, they follow Hamiltonians that may be superpositions over all $4^N$ of the generalized Pauli basis directions, $H = \sum_I h_I \sigma_I$. A geodesic that moves diagonally will be much shorter than a gate-path that follows a Manhattan grid. Since the grid is $4^N$-dimensional, the Pythagorean penalty could be as large as $2^N$. This forces us to cut down the number of directions. We divide directions into `expensive' and `cheap'. For the expensive directions (that cost more than some value $\overline{\mathcal{I}}$), we simply prune them out of the Hamiltonian. Excision introduces error, but at fixed geodesic-length $L$ and when $\overline{\mathcal{I}}$ is sufficiently large the error must be small, since the Hamiltonian cannot have a large component in expensive directions without the geodesic becoming long. For the cheap directions we simply eat the squareroot slow down (and  budget for the error caused by the monomials not commuting). Next we optimize the dividing line between cheap and expensive, $\overline{\mathcal{I}}$, to give the tightest bounds. This is altogether a more delicate balancing act than what Nielsen \emph{et al.} had to do for the cliff metric \cite{Nielsen2}, since in that case there was a small number of cheap directions and the rest were all exponentially expensive, whereas our more gradual  schedules must contend with the proliferation of mid-market penalty factors that are both moderately inexpensive and vexatiously numerous.

\subsection{Complexity geometries and BQP}
The complexity class `bounded-error quantum polynomial time' (BQP) is the set of decision problems that can be solved in polynomial time on a quantum computer, with low probability of error for every input. This is known to be equivalent to the set of decision problems that can be solved using a uniform family of quantum circuits with polynomially many gates (again requiring low probability of error for every input). 

Let's collect three relevant results. First, we saw in Eq.~\ref{eq:clifflessthanpigates} that complexity geometries can efficiently simulate circuits exactly (so long as $\mathcal{I}_2$ is at-worst polynomially large). Second, we saw that circuits can efficiently simulate complexity geometries that satisfy Eqs.~\ref{sufficientcondition2} or \ref{sufficientcondition3}, 
so long as we tolerate small operator-norm errors. Finally  in Sec.~\ref{sec:statecomplexity} we saw that tolerating small operator-norm errors will lead to low probability of error even for the worst-case inputs. 
 
These results tell us that the ability to enact every polynomial-length path through a complexity geometry that satisfies Eq.~\ref{sufficientcondition0} and Eqs.~\ref{sufficientcondition2} or \ref{sufficientcondition3} confers the ability to solve a class of decision problems with high probability for every input if and only if those decision problems are in BQP.   Any complexity geometry that satisfies Eq.~\ref{sufficientcondition0} and  Eqs.~\ref{sufficientcondition2} or \ref{sufficientcondition3} thus gives an alternative definition of BQP-completeness.

(We will also need to dispense with two technicalities. First, the definition of BQP  requires not only that there be a family of circuits that implements the unitary, but also that the family of circuits be `uniform', meaning that there be an efficient classical algorithm that compiles the circuit. This condition is satisfied by our simulation algorithm: the procedure in Sec.~\ref{sec:sec3} not only proved the existence of a circuit that approximates a given geodesic through the complexity geometry, it also gave an explicit polynomial-time recipe for constructing it. Second, the definition of BQP allows us to act the circuit not only upon the $N$-qubit input state, but also on an additional set of `ancilla' qubits. These qubits are all initialized at zero, and can be thought of as a quantum workspace.  One might worry that the ability to use these ancilla qubits makes the circuit definition \emph{more} powerful than the complexity geometry definition, since the circuit now describes a path through the space of unitaries on the extended Hilbert space that also includes the  ancilla qubits.  However, no polynomial-size circuit can reach more than a polynomial number of ancillas, so we can use the same method that led to Eq.~\ref{eq:clifflessthanpigates} to construct a polynomial-length complexity geometry path through the extended Hilbert space that exactly emulates the circuit.) 

\subsection{The role of the binomial metric}  \label{subsec:vastrightinvariantconspiracy}
 Recall that the binomial schedule, Eq.~\ref{eq:examplemetric2}, is given by 
\begin{equation}
\mathcal{I}_k =  \left(  \mathcal{N}_{k}\right)^\alpha  \equiv \left(  {N \choose k} 3^{k} \right)^\alpha \ .  
\end{equation}
For $\alpha =1$ the penalty factor of a $k$-local basis direction $\sigma_I$ is equal to the total number of $k$-local basis directions, $\mathcal{N}_k$. The $\alpha = 1$ binomial metric has shown up in three of our calculations, in each case playing a special role. Let's review them here.

The first place the binomial metric plays a special role is in the sufficient condition Eq.~\ref{sufficientcondition2} for polynomial equivalence between the complexity metric and the gate definition of complexity with operator-norm error. To control the operator-norm pruning error we need to be able to make $\sum_I \mathcal{I}(\sigma_I)^{-1}$ less than any polynomially small value by omitting from the sum at most polynomially many basis directions. As discussed around Eq.~\ref{eq:alphaisoneisspecial}, this leads to a critical value $\alpha =1$. For $\alpha > 1$ the metric satisfies the sufficient condition; for $\alpha \leq 1$ there is no guarantee that there are not polynomially long paths to unitaries that cannot be operator-norm approximated without superpolynomially many gates. (As discussed in Sec.~\ref{subsubsec:imcrementalimprovements}, it seems likely that by making use of the structure of the commutation relations we could also show the $\alpha = 1$ metric is in the equivalence class.) 

The second place the binomial metric plays a special role is in Eq.~\ref{eq:sufficientconditionoperatoriscomplex}. This is the sufficient condition that guarantees that being close in the complexity geometry implies being close in the operator norm. Again the quantity $\sum_I \mathcal{I}(\sigma_I)^{-1}$ is important, and for essentially the same reason, though this time we only require that $\sum_I \mathcal{I}(\sigma_I)^{-1}$ be at most polynomially large. The binomial metric with $\alpha \geq 1$ satisfies the sufficient condition, whereas for $\alpha <1$ there may be paths that are short in complexity distance but not short in operator-norm distance. 

The final place the $\alpha = 1$ binomial metric plays a special role is in Ref.~\cite{Brown:2021euk}. In that paper I lowerbounded the  diameter of the complexity geometry. (The lowerbound was proved using a technique from differential geometry, and \emph{not} using the quantum-simulation methods we  deployed in Sec.~\ref{subsec:implicationsfordiameter} of this paper to derive a weaker bound on the same quantity.)  
According to the lowerbound in  \cite{Brown:2021euk}, if $\mathcal{N}$ of the directions are cheap and the rest are very expensive, the total Killing distance required to hit every unitary scales like $2^N \mathcal{N}^{-1/2}$. This means the complexity distance needed to hit every unitary using only the exactly $k$-local directions is lowerbounded by $2^N  \sqrt{\mathcal{I}_k/ \mathcal{N}_k}$. If this lowerbound is tight, then the $\mathcal{I}_k = \mathcal{N}_k$ metric 
 is the critical `load balanced' schedule for which the cost to synthesize a typical unitary using the exactly $k$-local directions is independent of $k$.

\subsection{Equivalence class of complexity geometries}  \label{subsec:geometricperspectives}

Let's take a geometrical perspective. From the point of view of geometry, the relevant result in this paper is not that right-invariant geometries that satisfy certain sufficient conditions are polynomial equivalent to the gate definition of complexity, it's rather that these right-invariant geometries are polynomially equivalent to each other.

In a previous paper \cite{Brown:2021rmz}, my collaborators and I conjectured that the long-distance geometry of right-invariant metrics should be markedly insensitive to short-distance parameters. Even if two spaces look very different at short scales, they may nevertheless give rise at longer scales to the same emergent effective geometry. We made an analogy with the UV/IR decoupling that happens in the theory of renormalization in quantum field theory. In our paper we gave simple examples of this phenomenon at work, and made quantitative conjectures for the large-separation distance function for high-dimensional right-invariant metrics on Lie groups. The results of this paper provide further support for these conjectures. The short-distance properties of the metric, like the curvatures and the small-separation distance functions, are given directly by the penalty factors. The long-distance properties of the metric like the large-separation distance functions are also formally determined by the penalty factors, but in a more convoluted fashion that shields the `IR' from the details of the `UV'. What we have shown in this paper is that even when two metrics differ exponentially in their short-distance properties (e.g.~the binomial metric and the cliff metric differ exponentially in their assignment of penalty factors to moderate-weight directions) they may nevertheless be in the same polynomial equivalence class for approximate synthesis at long distances.

\subsection{Prospects for improving bounds} \label{subsec:prospectsforimproving}
Let's consider the prospects for tightening the bounds we have derived in this paper. 

\subsubsection{Incremental improvements} \label{subsubsec:imcrementalimprovements}
The most straightforward and least consequential improvements to Eqs.~\ref{eq:boxedgettingFrobeniusclose}, \ref{eq:boxedgettingoperatorclose}, \& \ref{eq:boxedgettingoperatorclose2} would be to try to improve the exponents of $L$, $s($error$)$, $||\textrm{error}||_\textrm{op}$, and $\mathcal{N}_{\overline{\mathcal{I}}}$, while still tolerating bi-invariant error and while still settling for polynomial rather than linear equivalence. One approach might be to replace the first-order Trotterization used in Sec.~\ref{subsec:trottererror} with a higher-order Suzuki-Trotter formula; another might be to deploy some more state-of-the-art quantum simulation techniques such as those in Refs.~\cite{posttrotter1,posttrotter2,posttrotter3,posttrotter4,posttrotter5}. 

However, even within the confines of first-order Trotterization  there is reason to be hopeful that the $\sqrt{\mathcal{N}}$ scaling of the normalized Frobenius norm of the Trotter error in Eq.~\ref{eq:referbacktothissquareroottocastdoubtonit} might be improved. 
As discussed in Appendix~\ref{appendix:howdoesTrotterscalewithN}, the $\sqrt{\mathcal{N}}$ scaling  of the  O($\delta^2$) error term is caused by the possibility of constructive interference between the various commutators. But using our freedom to choose the order of the factors within the Trotter product, we can arrange for the interference to be on-net destructive. If this result---laid out in Appendix~\ref{appendix:howdoesTrotterscalewithN}---could be extended to all orders in $\delta$, then the bound on the per-segment Trotter error described in Eq.~\ref{eq:referbacktothissquareroottocastdoubtonit} would improve from  $\mathcal{N}^{1/2} \delta^2$ to 
\begin{equation}
\textrm{conjecture}: \ \  \ \  \  s(\textrm{trotter error}) \ \ \lsim \ \ \mathcal{N}^0 \delta^2 \ . \ \ \ \ \ \ \ \ \ \ \ 
\end{equation}
This in turn would improve the upperbound in Eq.~\ref{eq:boxedgettingFrobeniusclose} to $ \mathcal{C}_\textrm{gates} \ \lsim \    \mathcal{N}_{\overline{\mathcal{I}}}  L^2/ s(\textrm{error})$. This upperbound would be particularly intriguing because it would imply that the lowerbound on the diameter of the complexity geometry  we could derive using the technique of Sec.~\ref{subsec:implicationsfordiameter} would be identical to that proved using a completely different technique in \cite{Brown:2021euk}. 

Another set of improvements would be to make use of the structure of the commutation relations amongst the $k$-local generalized Pauli's. In our analysis so far, we more-or-less made worst-case assumptions about the commutation relations. But for penalty schedules in which the cheap directions are the $k$-local directions with small $k$, we can make use of the fact that the commutation relations are highly structured. (As an example of the structure, the number of independent commutators of the $\mathcal{N}_k$ $k$-local directions is much smaller than $\mathcal{N}_k^2$). This should allow us to improve the bounds both for getting Killing close and for getting operator-norm close. For example, consider the derivation of the bound on the operator-norm pruning error in Sec.~\ref{subsubsec:operatornormpruning}. We used the inequality $||\sum_I h_I \sigma_I ||_\textrm{op} \leq  \sum_I || h_I \sigma_I ||_\textrm{op}$. But that inequality is generally not tight. As discussed in Sec.~\ref{subsubsec:operatornormerror} the operator norm is often smaller than the sum of its parts. By making use of the structure of the commutation relations between $k$-local operators we could sharpen our bound on the critical schedule in Sec.~\ref{subsubsec:binomialmetricoperatorclose} to be less expensive than $\mathcal{I}_k = \mathcal{N}_k \equiv {N \choose k} 3^k$. (However, there is no prospect that this could show the critical metric to be any less punitive than $\mathcal{I}_k =  {N \choose k}$ because, as discussed in Sec.~\ref{subsubsec:operatornormerror}, the inequality $||\sum_I h_I \sigma_I ||_\textrm{op}  \leq  \sum_I || h_I \sigma_I ||_\textrm{op}$ \emph{is} saturated if we only include the ${N \choose k}$ commuting $k$-local generalized Pauli's that are exclusively composed of $\mathds{1}$s and $\sigma_z$s (with no $\sigma_x$s or $\sigma_y$s).) 

Relatedly, we can relax our condition Eq.~\ref{sufficientcondition0} that, to belong to the equivalence class, penalty schedules must assign two-local directions modest penalty factors. The motivation for this condition was to guarantee the `only if' direction of the equivalence:  to show that complexity geometries can easily emulate gates. But it's overkill. It is sufficient to require that there is \emph{some} set of cheap generalized Pauli's whose nested commutators generate the entire algebra. With that set, we can use the technique of Sec.~\ref{sec:makingmonomials} to make any gate. What this means is that we can permute the penalty factors amongst the generalized Pauli's and, subject only to the condition that there always be a cheap subalgebra whose nested commutators generate the entire algebra, the metric will remain in the same polynomial equivalence class. Such a permutation does \emph{not} need to respect any notion of $k$-locality (although it must be a permutation rather than a more general rotation---it does need to keep the penalty matrix diagonal in the generalized-Pauli basis, see below). As an illustration of this,  consider that there are actually \emph{two} natural tensor decompositions of U$(2^N)$. One is the decomposition into $N$ qubits; the other is the decomposition into $2N$ Majoranas. These two decompositions are related by the Jordan-Wigner transformation. The Jordan-Wigner transformation is a permutation that maps diagonal penalty matrices to diagonal penalty matrices, but does not respect  $k$-locality, so that an operator that touches only a few qubits may touch many Majoranas, or vice versa.  Just as we defined e.g.~an `exponential metric' with respect to qubits, which punishes generalized Pauli's proportional to the exponential of the number of qubits they touch, so too could we define a Majorana version of the `exponential metric' that punishes generalized Pauli's proportional to the exponential of the number of Majoranas. A consequence of our analysis is that these two metrics are in the same quasi-polynomial equivalence class---subject to tolerating Killing error---despite rendering completely different directions cheap. Similarly, when we tolerate operator-norm error, a parallel argument to the one that leads to Eq.~\ref{eq:alphaisoneisspecial} tells us that the critical Majorana metric is $\mathcal{I}_k = {2N \choose k}$ and that any penalty schedule parametrically more expensive than this will be in the same polynomial equivalence class as quantum circuits.

Finally, let us comment on the restriction that the penalty matrix $\mathcal{I}_{IJ}$ must be diagonal in the generalized-Pauli basis. There is no prospect of being able to fully relax this condition: we will not be able to make a useful sufficient condition for polynomial equivalence that is a function only of the spectrum of  $\mathcal{I}_{IJ}$, with no reference to what basis $\mathcal{I}_{IJ}$ is diagonal in. To see the obstruction, consider a penalty matrix that, while generally having large eigenvalues, makes it cheap to move in a few Haar-random directions. These cheap directions will be approximately equal superpositions over all $4^N$ of the generalized Pauli's. No known simulation technique makes it cheap to approximate motion in a Haar-random direction using gates, and a counting argument tells us that no simulation technique could possibly work in the typical case (and vice-versa: the Haar-random directions will not themselves be able to easily simulate gates). It's not that the `off-diagonal' penalty matrix gives a more powerful model of computation, it's rather that the power of the two models of computation are simply incommensurate.

\subsubsection{Errors as measured in complexity geometry  do not compose}

In the bounds proved in this paper, we measured the length of the path $L$ in the complexity geometry metric, but measured the error in a different metric (operator norm, Frobenius norm, or Killing). One might aspire to prove a stronger theorem in which we measure both the length of the path \emph{and} the targeted error in the complexity geometry. Unfortunately, I will now show that we cannot prove such a bound with a simple upgrade of the method of this paper nor indeed using any strategy of divide-and-conquer.

In a divide-and-conquer strategy (such as the one we deployed in Sec.~\ref{sec:sec3}) we break the path into many small segments, bound the error of each small segment, and then bound the total error by adding up the segment errors. The `bound the segment error' step is not the problem: for gradual penalty schedules (for which the penalty factor only gets large at large weight), we can bound the complexity-geometry error of each segment by noticing that the Taylor expansion for $\delta$ (see e.g. Eq.~\ref{eq:errorintermediatepreamble}) is also an expansion in the weight of the nested commutator. Instead the problem is the `add up the segment errors' step. Since the complexity geometry  is only right-invariant and not also left-invariant, there is no guarantee that $L(U_1 U_{2},U_{1'} U_{2'}) \leq L( U_{1}, U_{1'}) + L(U_{2}, U_{2'})$. The divide-and-conquer strategy will not work because composing two steps both of which have small error may nevertheless lead to a combined path that is unacceptably erroneous.

\subsubsection{Beyond divide-and-conquer}

Finally, and most ambitiously, we might aspire to completely eliminate the need to tolerate error, and perhaps even establish a linear bound between some definitions of complexity. Certainly the existence of a robust linear relationship between a set of definitions of complexity would be welcome news for the holographic complexity program, which via the AdS/CFT correspondence ascribes physical meaning---the size of the dual wormhole---to the numerical value of the complexity \cite{Susskind:2014rva,Stanford:2014jda,Brown:2015bva,Brown:2015lvg}. The more robust this numerical value can be shown to be, the more robust the foundation of holographic complexity. If this is possible, it will require a radical departure from the methods of this paper, and require transcending not only Suzuki-Trotter but also the entire strategy of divide-and-conquer. 

This is probably not possible for gates. 
We have seen that it is challenging to follow complexity geometry paths with gates, 
because gates move through the unitary group only in clumsy zig-zagging quantum leaps. With such poor fine-motor skills,  gates probably cannot yield bounds vastly superior to the ones we have already derived, so   something qualitatively similar to Eqs.~\ref{eq:boxedgettingFrobeniusclose}, \ref{eq:boxedgettingoperatorclose}, \& \ref{eq:boxedgettingoperatorclose2} is about as good as it gets.

Where the situation is more promising is for one complexity geometry simulating another. As recounted in Sec.~\ref{subsec:geometricperspectives}, in a previous paper \cite{Brown:2021euk} my collaborators and I conjectured that one right-invariant metric on a high-dimensional Lie group can simulate another with much greater accuracy, and much lower overhead, than the analysis of Sec.~\ref{sec:sec3} would  predict. As well as some suggestive examples, we also gave a couple of concrete pieces of evidence that such an improved simulation strategy should exist. 

First, the `ball-box theorem' of Gromov \cite{gromov1996carnot} says, in our language, that for any right-invariant metric on the unitary group,  so long as the nested commutators of the cheap directions generate the entire algebra, then even if the other directions are infinitely expensive we can still economically fill out the entire neighborhood of a point, hitting anywhere within the neighborhood with zero error. This fact is completely invisible at any finite order in the Suzuki-Trotter expansion. 

Second, in our paper we carefully examined the complexity geometry of a single qubit \cite{Brown:2019whu,Brown:2021euk}---the squashed three-sphere---and found not only that we could hit every point with zero error, but also that the incremental cost of eliminating error got smaller and smaller the farther away we were aiming. This behavior makes no sense in the context of a Suzuki-Trotter expansion, or more generally in the context of any strategy of divide-and-conquer that involves dividing the path into small segments, simulating each in turn, and then adding up the errors from each segment. Any divide-and-conquer strategy will predict that---just as we see with the formulas we derived in Sec.~\ref{sec:sec3}---the greater the separation the harder it is to control the error. But it makes perfect sense in the context of a conjectured more holistic compilation strategy that from the beginning sees all the way to the very end of the simulation---not just the end of the segment---and is able to plan far enough in advance to make micro-adjustments to the trajectory en route that leverage the greater lever-arm provided by the greater distance to the destination. Such a global strategy would make essential use of the `wiggle room' provided by the fact that Riemannian paths can change direction without penalty, and so there is an infinite dimensionality of small deformations to the path we can make as we seek to eliminate error. 

This phenomenon can be expected to be particularly pronounced for high-dimensional spaces. While high-dimensional spaces have a greater number of directions in which to err, they permit a greater number of small deformations that can be deployed to fix those errors, and in Ref.~\cite{Brown:2021euk} we argued that as a statistical matter the latter consideration should dominate. (We saw a foreshadowing of this phenomenon in the discrete context of Sec.~\ref{subsubsec:imcrementalimprovements}, where we were able to harness the exponential number of degenerate deformations of the first-order Trotter formula to force the errors to destructively interfere.) 

There are a number of tantalizing hints that point to the existence of a holistic simulation strategy that is qualitatively superior to any currently known. Unfortunately, all these hints are non-constructive. The path may exist, but I have no idea how to find it.

\subsection{Curvature and complexity} 
Years ago, Nielsen laid out a program \cite{Nielsen1} to use the tools of differential geometry to prove complexity lowerbounds. His idea was that by recasting quantum complexity in geometric terms, we might harness the great machinery of that branch of mathematics, and bypass some of the barriers that impede more conventional approaches. Unfortunately, progress has been slow. Let me finish this paper with some speculation on why progress has been slow, and on why there is reason to hope it might not remain slow.  An essential role will be played by Riemannian 
curvature.

The complexity geometry is curved. The largest sectional curvatures are when two cheap directions $\sigma_I$ and $\sigma_J$ commute to an expensive direction, giving   approximately \cite{Milnor} 
\begin{equation}
\mathcal{R}_{IJ}^{\ \ JI} \ \sim \  - \frac{\mathcal{I}([\sigma_I,\sigma_J]) }{ \mathcal{I}(\sigma_I)\mathcal{I}(\sigma_J)} \ .
\end{equation} 
This tells us that the cliff metric, Eq.~\ref{eq:examplemetric1}, is very curved indeed. Two (cheap) 2-local directions  that commute to a (maximally expensive) 3-local direction give a sectional curvature that is exponentially big. 
The cliff metric is so highly curved that while it is still technically Riemannian  it is difficult to exploit its Riemannian-ness.  Perhaps this explains why the Nielsen program has made little progress. 
The tools of differential geometry have a hard time getting purchase when the curvature is exponential.

 But not all complexity geometries have this problem. Less abrupt penalty schedules like the exponential schedule and the binomial schedule---for which the penalty factors grow  slowly with weight---have modest curvature  \cite{secondlaw}. The commutator of a $k$-local direction and an $m$-local direction can never be more than ($k$+$m$-$1$)-local, so any penalty schedule that has $\mathcal{I}_{k+m-1}  \lsim \, \mathcal{I}_{k} \mathcal{I}_{m}$ will have modest sectional curvature for every section.

The results of this paper enlarge the equivalence class of complexity geometries known to be polynomially equivalent to the gate-counting definition of complexity. Previously, the class was only known to contain geometries---like the cliff metric---that have extremely high curvature. The enlarged class now includes geometries---like the binomial metric---that have modest curvature. Perhaps the tools of differential geometry that failed when applied to the cliff metric might now meet with more success.\footnote{I took a step in this direction in \cite{Brown:2021euk} when I used a tool from differential geometry---the Bishop-Gromov theorem---to harness the more gentle Ricci curvature of some less abrupt penalty schedules to prove a complexity lowerbound, though in that case not  making use of the polynomial equivalence proved in this paper.} 

\section*{Acknowledgments }
Thank you to Michael Freedman, Henry Lin, and Leonard Susskind for collaboration and discussions. And thank you to Joseph Malkoun and Victor Ramos Puigvert for catching typos in version one of this manuscript.

\appendix

\section{Proving lemmas} 

In this appendix we'll prove the  lemmas we relied on in the main text.

\subsection{Upperbounding $s$ in terms of the Frobenius distance} \label{sec:Frobboundsinnerproduct}
The Killing distance $s$ is always greater than the normalized Frobenius-norm distance. 
Let's show that even though $s$ is bigger, it's not that much bigger. We will prove Eq.~\ref{eq:tobeprovedinappendix1}.  

Every unitary can be written in the form $e^{i H t}$ for some time-independent $H$. In fact there are many such Hamiltonians because the matrix logarithm is not unique. These time-independent Hamiltonians generate all the geodesics of the Killing geometry (see Eq.~\ref{eq:ArnoldKhesin}), and the shortest geodesic will be generated by the $H$ that reaches the target with the smallest value of $\sqrt{\overline{\textrm{Tr}}H^2}  t$. If the eigenvalues of $H t$ for this shortest geodesic are $\{ \lambda_1, \lambda_2, \ldots \lambda_{2^{N}}\}$, then the Killing distance is 
\begin{equation}
s(\mathds{1},e^{iHt}) = \sqrt{\overline{\textrm{Tr}}[H^2]} t = 2^{-N/2} \sqrt{ \sum_{i=1}^{2^N} \lambda_i^{\, 2}}    \ . 
\end{equation} 
Since $H$ generates the \emph{minimal} geodesic, we must have 
\begin{equation}
    |\lambda_i| \leq \pi \ ; 
\end{equation}
for larger values of $\lambda_i$ we have gone the long way round the U(1) circle for that eigenvalue (see Fig.~\ref{fig:chordorcircumference}) and we can shorten the path by instead going the short way. This confirms that the most distant unitary from the identity $\mathds{1}$ is the antipodal $- \mathds{1}$, which has $\lambda_i = \pi$ for all eigenvalues and sits at a distance $s = \pi$.

Now let's consider the normalized Frobenius-norm distance, defined in Sec.~\ref{subsec:frobeniusnormdefinition}, 
\begin{equation}
||\mathds{1} - e^{iHt} ||_{\overline{F} } = \sqrt{2 - 2 \, \overline{\textrm{Tr}} \cos H t} =  \sqrt{ 2^{-N} \sum_{i=1}^{2^N} (2 - 2\cos \lambda_i) } = 2^{-N/2} \sqrt{ \sum_{i=1}^{2^N} \Bigl( 2 \sin \frac{\lambda_i}{2} \Bigl)^2} \ . 
\end{equation}
We wish to bound the ratio between the Killing distance and the normalized Frobenius-norm distance, 
\begin{equation}
    \frac{ s(\mathds{1},e^{iHt})}{ ||\mathds{1} - e^{iHt} ||_{\overline{F} } } =  \sqrt{\frac{ \sum_i \lambda_i^2 } {\sum_i \bigl( 2 \sin \frac{\lambda_i}{2} \bigl)^2 } } \ . 
\end{equation}
Since $|\lambda_i| \leq \pi$ implies $1< \frac{\lambda_i}{2 \sin \frac{\lambda_i}{2} }  < \frac{\pi}{2}$, this means that 
\begin{equation}
  ||\mathds{1} - e^{iHt} ||_{\overline{F}} \  \leq \ s(\mathds{1},e^{iHt}) \ \leq \ \frac{\pi}{2}  ||\mathds{1} - e^{iHt} ||_{\overline{F} } \ . \label{eq:finalrelationshipbetweenFbarands}
\end{equation}
Bi-invariance then implies Eq.~\ref{eq:tobeprovedinappendix1}.

\subsection{Bounding the error from pruning} 
Approximation 1, Eq.~\ref{eq:approximation1}, was to drop all terms that have a penalty factor greater than some critical value $\overline{\mathcal{I}}$,
\begin{equation}
H(t) = \sum_I h_I(t) \sigma_I \ \rightarrow \ H_P(t) \equiv \sum_{I \textrm{ with }  \mathcal{I}(\sigma_I) < \overline{\mathcal{I}}}  h_I(t) \sigma_I \ . 
\end{equation} 
This will introduce error, which we will now bound.  Eq.~\ref{eq:normpathlessthanhamiltonian} says that for any bi-invariant norm 
\begin{equation}
\bigl|\bigl|  \mathcal{P} e^{i \int H(t) dt} -  \mathcal{P} e^{i \int H_P(t) dt} \bigl|\bigl|   \leq   \int dt || H(t) - H_P(t) || =   \int dt \, \bigl| \bigl| \hspace{-2mm} \sum_{\mathcal{I}(\sigma_I) > \overline{\mathcal{I}}}  h_I(t) \sigma_I  \bigl| \bigl|    \ .  \label{eq:integralofnorm}
\end{equation}
To bound the right-hand side of this expression, we will use Eq.~\ref{eq:conservedhardness}, 
\begin{equation}
\Gamma^2 = \sum_I \mathcal{I}(\sigma_I) h_I(t)^2  \ . \label{eq:conservedhardness2}
\end{equation}

\subsubsection{Pruning error for Frobenius norm}
The normalized-Frobenius pruning error is upperbounded by Eq.~\ref{eq:integralofnorm} as 
\begin{equation}
|| \mathcal{P} e^{i \int H(t) dt} -  \mathcal{P} e^{i \int H_P(t) dt} ||_{\overline{F}}  \ \leq \  \int dt \sqrt{ \sum_{\mathcal{I}(\sigma_I) > \overline{\mathcal{I}}}  h_I(t)^2}  \ . 
\end{equation}
We can us Eq.~\ref{eq:conservedhardness2} to write 
\begin{equation}
\Gamma^2 = \sum_I \mathcal{I}(\sigma_I) h_I(t)^2 \geq   \overline{ \mathcal{I}} \sum_{ \mathcal{I}(\sigma_I) > \overline{\mathcal{I}}} h_I(t)^2  \ \  \rightarrow \ \  \sum_{ \mathcal{I}(\sigma_I) > \overline{\mathcal{I}}} h_I(t)^2  \leq \frac{\Gamma^2}{ \overline{\mathcal{I}}} \ . 
\end{equation}
Combining these equations gives 
\begin{equation}
|| \mathcal{P} e^{i \int H(t) dt} -  \mathcal{P} e^{i \int H_P(t) dt} ||_{\overline{F}}   \leq \int dt \frac{\Gamma}{\sqrt{\overline{ \mathcal{I}}}}  =\frac{L }{\sqrt{\overline{ \mathcal{I}}}} \ . \label{eq:boundingpruningfrobeniusfinal}
\end{equation} 
Alternatively we could have derived the same inequality by applying Eq.~\ref{eq:RiemannianbiggerthanFrobenius} to Eq.~\ref{eq:innerproducterrorexcision}.

\subsubsection{Pruning error for operator norm} \label{sec:operrorpruningformula} 
There are two different upperbounds we can place on the operator-norm pruning error, for the same reason there are two upperbounds in Eq.~\ref{eq:norminequalitieswithboth}.\\ 

\noindent {\bf First bound.} The first bound come from applying subadditivity to Eq.~\ref{eq:integralofnorm} to get 
\begin{equation}
\bigl|\bigl| \mathcal{P} e^{i \int H(t) dt} -  \mathcal{P} e^{i \int H_P(t) dt} \bigl|\bigl|_{\textrm{op} }  \ \leq \  \int dt \sum_{\mathcal{I}(\sigma_I) > \overline{\mathcal{I}}}    || h_I(t) \sigma_I  ||  \ = \  \int dt  \sum_{\mathcal{I}(\sigma_I) > \overline{\mathcal{I}}}  |h_I(t)|  \ .  \label{eq:operatorpruning1}
\end{equation}
Let's upperbound the right-hand side using Eq.~\ref{eq:conservedhardness2}. To maximize $\sum_I |h_I|$ at fixed $\sum_I   \mathcal{I}(\sigma_I)  h_I^2$, the worst-case scenario is 
\begin{equation}
\textrm{maximize $\sum_I |h_I|$ at fixed $\Gamma$}:  \ \ \ \ \ h_I = \frac{ {c}}{\mathcal{I}(\sigma_I)} , \ \ \ \ \ \ \ \ \ \ \ \hspace{2cm} 
\end{equation} 
for some Lagrange multiplier $c$. This worst-case-scenario gives 
\begin{equation}
h_I = \frac{ {c}}{\mathcal{I}(\sigma_I)} \ \ \rightarrow \ \   \sum_{\mathcal{I}(\sigma_I) > \overline{\mathcal{I}}}  |h_I(t)|  = c \sum_{\mathcal{I}(\sigma_I) > \overline{\mathcal{I}}}  \frac{1}{\mathcal{I}(\sigma_I)} = c^{-1} \sum_{\mathcal{I}(\sigma_I) > \overline{\mathcal{I}}}  \mathcal{I}(\sigma_I) h_I(t)^2   \ \leq \ c^{-1} \Gamma^2  \ . 
\end{equation}
Eliminating $c$, this implies 
\begin{equation}
\sum_{\mathcal{I}(\sigma_I) > \overline{\mathcal{I}}}  |h_I(t)|   \leq \Gamma \sqrt{ \sum_{\mathcal{I}(\sigma_I) > \overline{\mathcal{I}}}  \frac{1}{\mathcal{I}(\sigma_I)} }  \ . 
\end{equation}
Combining this with 
Eq.~\ref{eq:operatorpruning1}
gives 
\begin{equation}
\bigl|\bigl| \mathcal{P} e^{i \int H(t) dt} -  \mathcal{P} e^{i \int H_P(t) dt} \bigl|\bigl|_{\textrm{op} }  \ \leq  \  L \sqrt{ \sum_{\mathcal{I}(\sigma_I) > \overline{\mathcal{I}}}  \frac{1}{\mathcal{I}(\sigma_I)} }  \ . \label{eq:operrorpruningformula}
\end{equation} \\

\noindent {\bf Second bound.} The second bound comes from applying Eq.~\ref{eq:norminequalitieswith2totheN} to Eq.~\ref{eq:boundingpruningfrobeniusfinal} to give 
\begin{equation}
\bigl|\bigl| \mathcal{P} e^{i \int H(t) dt} -  \mathcal{P} e^{i \int H_P(t) dt} \bigl|\bigl|_{\textrm{op} }  \ \leq  \ \frac{L\, 2^{N/2}}{\sqrt{ \overline{ \mathcal{I}}}}  \ . 
\end{equation}

\subsection{Bounding the error from averaging}
Approximation 2 was to average the Hamiltonian within each segment, Eq.~\ref{eq:approximation2}, replacing $H_P(t)$ with ${H_\textrm{av.}}  \equiv \delta^{-1} \int_{0}^{ \delta} H_P(t) dt$. 
We can upperbound the norm error this introduces using the Dyson expansion
\begin{eqnarray}
\textrm{err}_\textrm{av} & \equiv&  \Bigl|\Bigl|     \exp\Bigl[ i \int_{0}^{ \delta} H_P(t) dt \Bigl]  -  \mathcal{P} \exp\Bigl[ i \int_{0}^{ \delta} H_P(t) dt \Bigl]  \Bigl|\Bigl|   \\
& = & \Bigl|\Bigl|   \ \sum_{m=2}^\infty \frac{(- i {H_\textrm{av.}} \delta)^m}{m!}  - \sum_{m = 2}^\infty (-i)^m \int_0^\delta dt_1  \int_0^{t_1} dt_2 \ldots \int_0^{t_{m-1}} dt_m H_P(t_1) \ldots H_P(t_m) \Bigl|\Bigl| \nonumber \\
&  \leq &  \sum_{m=2}^\infty \frac{||(- i {H_\textrm{av.}} \delta)^m||}{m!}  + \sum_{m = 2}^\infty  \int_0^\delta dt_1   \ldots \int_0^{t_{m-1}} dt_m ||H_P(t_1) \ldots H_P(t_m) || \ . \label{eq:errorintermediatepreamble} 
\end{eqnarray}

\subsubsection{Averaging error for Killing distance} \label{subsec:averagingerrorfrobeniusappendix}
It is a general theorem that, for any $A$ and $B$, we have $||AB||_F \leq  ||A||_F ||B||_\textrm{op}$ and therefore 
\begin{eqnarray}
||AB||_{\overline{F}} & \leq & ||A||_{\overline{F}} ||B||_\textrm{op}  \ .
\end{eqnarray}
By induction, and coupled with Eq.~\ref{eq:norminequalitieswithN}, this implies that 
\begin{equation}
||H_P(t_1) H_P(t_2) \ldots H_P(t_m)||_{\overline{F}}   \ \leq \ \mathcal{N}^{ \frac{m-1}{2}}  ||H_P(t_1)||_{\overline{F}}  ||H_P(t_2)||_{\overline{F}}  \ldots ||H_P(t_m)||_{\overline{F}}  \ . 
\end{equation} 
Since we normalized $t$ so that  $||H_P(t)||_{\overline{F}}  \leq 1$, this implies that 
\begin{equation}
 || \mathcal{P}  \exp\Bigl[ i \int_{0}^{ \delta} H_P(t) dt \Bigl]  -  \exp\Bigl[ i \int_{0}^{ \delta} H_P(t) dt \Bigl]  ||_{ \overline{F} }  \ \leq \  2 \sum_{m=2}^\infty \frac{ \mathcal{N}^{\frac{m-1}{2} }  \delta^m }{m!} 
\leq \frac{2 \left( e^{\sqrt{ \mathcal{N}} \delta} - 1 - \sqrt{ \mathcal{N}} \delta \right) }{ \sqrt{ \mathcal{N}}}  \ .  \nonumber 
\end{equation}
Noting that $1 < 2y^{-2} (e^y - 1 - y) < 2$ for $0<y<1$, this implies
\begin{equation}
\delta < \mathcal{N}^{-\frac{1}{2}}  \ \ \ \ \rightarrow \ \  \ \ || \mathcal{P}  \exp\Bigl[ i \int_{0}^{ \delta} H_P(t) dt \Bigl]  -  \exp\Bigl[ i \int_{0}^{ \delta} H_P(t) dt \Bigl]  ||_{ \overline{F} }  \ < \ 2 \sqrt{\mathcal{N}} \delta^2 \ . \ \ \ \ \ \ \ \ \  \label{eq:answerforaveragingfrobeniuserror}
 \end{equation}

 \vspace{2mm}
\noindent Using Eq.~\ref{eq:finalrelationshipbetweenFbarands}, we can turn this from a bound on the Frobenius-norm error to a bound on the Killing error, 
\begin{equation}
\delta < \mathcal{N}^{-\frac{1}{2}}  \ \ \ \ \rightarrow \ \  \ \ s(\mathcal{P}  \exp\Bigl[ i \int_{0}^{ \delta} H_P(t) dt \Bigl] ,  \exp\Bigl[ i \int_{0}^{ \delta} H_P(t) dt \Bigl]  )  \ < \ \pi \sqrt{\mathcal{N}} \delta^2 \ . \ \ \ \ \ \ \ \ \  \label{eq:answerforaveragingbiinvariantRiemannian}
 \end{equation}

\subsubsection{Averaging error for operator norm} \label{subsec:averagingerroroperatorappendix}
The operator norm is submultiplicative, so using Eq.~\ref{eq:norminequalitieswithN},  we have that Eq.~\ref{eq:errorintermediatepreamble} becomes 
\begin{eqnarray} 
\bigl|\bigl| \mathcal{P}  \exp\Bigl[ i \int_{0}^{ \delta} H(t) dt \Bigl]  -  \exp\Bigl[ i \int_{0}^{ \delta} H(t) dt \Bigl]  \bigl|\bigl|_{\textrm{op}} &  \leq &  2 \sum_{m=2}^\infty \frac{( ||{H}||_{\textrm{op}} \delta)^m}{m!} \\
& \leq & 2 \sum_{m=2}^\infty \frac{(\sqrt{\mathcal{N}} \,  \delta)^m}{m!} \\
& = & 2 \left( e^{\sqrt{\mathcal{N}} \delta} - 1 - \sqrt{\mathcal{N}} \delta \right) \ .
\end{eqnarray}
Noting once again that $1 < 2y^{-2} (e^y - 1 - y) < 2$ for $0<y<1$, this implies
\begin{equation}
 \delta < \mathcal{N}^{-1/2} \  \ \rightarrow \ \ \ \bigl|\bigl| \mathcal{P}  \exp\Bigl[ i \int_{0}^{ \delta} H(t) dt \Bigl]  -  \exp\Bigl[ i \int_{0}^{ \delta} H(t) dt \Bigl]  \bigl|\bigl|_{\textrm{op}} \ < \  2 \mathcal{N} \delta^2 \ . \label{eq:errorfromaveragingprecise} 
\end{equation}
We could also have derived the same formula by applying Eq.~\ref{eq:norminequalitieswithN} to Eq.~\ref{eq:answerforaveragingfrobeniuserror}.

\subsection{Bounding the error from Trotterization}

In this subsection we will be interested in bounding the norm of the Trotter error, 
\begin{eqnarray}
\textrm{err}_\textrm{trot.} & \equiv&   \Bigl| \Bigl|  e^{ i \int_{T}^{T + \delta} dt \sum_I h_I(t)  {\sigma}_I  } -   \prod_{I = 1}^\mathcal{N}  \left( e^{i   \int_{T}^{T + \delta} dt \, h_I(t)  {\sigma}_I }  \right)  \Bigl| \Bigl|  \\
& = & \Bigl| \Bigl| \frac{1}{2}  \sum_{I} \sum_{J<I}  h_I h_J [ \sigma_I,\sigma_J] \delta^2 + O(\delta^3)   \Bigl| \Bigl| \ . 
 \end{eqnarray}
 There are two complications in bounding this expression: we must ensure that at non-infinitesimal $\delta$, the higher-order terms do not make a dominant contribution; and we must deal with the possibility that the O$(\delta^2)$ terms might constructively interfere. These questions have received extensive study \cite{Berry1,Wecker2014,Childs2019,Kivlichan2019,Layden,Zhao:2021gtg,Chen2021}, but for our purposes---at the cost of not fully optimizing our  estimates---we will use only the result (quoted in e.g.~Eq.~2 of \cite{Layden}) that for any submultiplicative bi-invariant matrix norm, 
 \begin{equation}
 \textrm{err}_\textrm{trot.}  \leq \frac{1}{2} \delta^2 \sum_I^{\mathcal{N}} \Bigl| \Bigl| \sum_{J>I}^{\mathcal{N}} h_I h_J [\sigma_I,\sigma_J] \Bigl| \Bigl|  \ . \label{eq:Trottererror}
 \end{equation}

\subsubsection{Trotter error for Killing distance} \label{appendix:trottererrorfrobeniusnorm}
For a given $\sigma_I$,  the commutator $[\sigma_I,\sigma_J]$  will be orthogonal to the commutator $[\sigma_I,\sigma_{J'}]$ if $J \neq J'$, so there will be no interference in Eq.~\ref{eq:Trottererror} and the terms inside the norm add in quadrature, 
 \begin{equation}
 \textrm{err}_\textrm{trot.} \Bigl|_{F} \  \leq \  \frac{1}{2} \delta^2 \sum_I^{\mathcal{N}} \sqrt{ \sum_{J>I}^{\mathcal{N}} h_I^2 \, h_J^2 \,  || [\sigma_I,\sigma_J]||_{{F}}^2 } \ . 
 \end{equation}
Using that for the normalized Frobenius error $||[\sigma_I,\sigma_J]||_{\overline{F}} \leq 2$, together with Eq.~\ref{eq:norminequalitieswithN}, this gives 
 \begin{equation}
 \textrm{err}_\textrm{trot.} \Bigl|_{\overline{F}} \  \leq  \  \delta^2 \sum_I^{\mathcal{N}} |h_I| \sqrt{\overline{\textrm{Tr}} H_P^2}  \   \leq  \ \delta^2 \sqrt{ \mathcal{N}} \, \overline{\textrm{Tr}}H_P^2 \ \leq  
 \ \delta^2 \sqrt{ \mathcal{N}} \ . 
 \end{equation}

 \noindent Using Eq.~\ref{eq:finalrelationshipbetweenFbarands}, we can turn this from a bound on the normalized Frobenius error to a bound on the Killing error, 
\begin{equation}
s\Bigl(e^{ i \int_{T}^{T + \delta} dt \sum_I h_I(t)  {\sigma}_I  } ,  \prod_{I = 1}^\mathcal{N}   e^{i   \int_{T}^{T + \delta} dt \, h_I(t)  {\sigma}_I }  \Bigl)  \ \leq \ \frac{\pi}{2} \sqrt{\mathcal{N}} \delta^2 \ . \ \ \ \ \ \ \ \ \ \label{eq:trotterbiinvariantquotesappendix}
 \end{equation}

\subsubsection{Trotter error for operator norm} \label{appendix:trottererrorperatornorm}
The Trotter error in the operator norm is bounded by Eq.~\ref{eq:Trottererror} as 
 \begin{equation}
 \textrm{err}_\textrm{trot.} \Bigl|_\textrm{op} \  \leq \  \frac{1}{2} \delta^2 \sum_I^{\mathcal{N}} \sum_{J>I}^{\mathcal{N}}  \Bigl| \Bigl| h_I h_J [\sigma_I,\sigma_J] \Bigl| \Bigl|_\textrm{op} \leq \  \delta^2  \ \sum_I^{\mathcal{N}} |h_I|  \sum_{J>I}^{\mathcal{N}} |h_J| \ \leq \ \delta^2 { \mathcal{N}} \ . 
 \end{equation}

\section{How does Trotter error scale with $\mathcal{N}$?} \label{appendix:howdoesTrotterscalewithN}
Consider Trotterizing the time-independent Hamiltonian ${H}=\sum_I^{\mathcal{N}}  h_I {\sigma}_I$. How does  the Frobenius norm of the Trotter error scale with $\mathcal{N}$? The difference between the first-order Trotter product and the  Hamiltonian evolution we are aiming for is 
\begin{equation}
\exp \Bigl[  i  \sum_{I=1}^{\mathcal{N}}  h_I  {\sigma}_I \delta  \Bigl] -   \prod_{I=1}^\mathcal{N}  e^{i     h_I  {\sigma}_I \delta }   = \frac{1}{2}  \sum_{I= 1}^\mathcal{N} \sum_{J=1}^{I-1}  h_I h_J [ \sigma_I,\sigma_J] \delta^2 + O(\delta^3)  \ . \label{eq:needforlater2}
 \end{equation}
 Let's discuss the normalized Frobenius norm of the O$(\delta^2)$ term. The complication is going to be interference. Interference happens when more than one term in the sum gives rise to the same operator---when $[\sigma_I,\sigma_J]$ is not orthogonal to $[\sigma_K,\sigma_L]$. If there were no interference, then the terms in the sum would just add in quadrature. Using $||[ \sigma_I,\sigma_J] ||_{\overline{F}} \leq 2$, we would have an expression without multiplicative factors of $\mathcal{N}$, 
 \begin{equation}
     \textrm{if no interference: } \ \ \ \ \Bigl|\Bigl| \frac{1}{2} \sum_{I= 1}^\mathcal{N} \sum_{J=1}^{I-1}  h_I h_J [ \sigma_I,\sigma_J] \Bigl|\Bigl|_{\overline{F}}^{\ 2}  \ \leq \  { \sum_{I= 1}^\mathcal{N} \sum_{J=1}^{I-1}  h_I^2 h_J^2 } \ \leq \ \frac{1}{2} (\overline{\textrm{Tr}} H^2)^2 \ . \label{eq:ifnointerference} 
 \end{equation}
When there is interference, rather than adding in quadrature the terms may just add (for constructive interference) or cancel (for destructive interference). If the number of different terms that give rise to the same commutator is $n$, and if all those terms constructively interfere, then  this could multiply the Frobenius norm by a factor of as much as $\sqrt{n}$.  How big can $n$ be? There are about $\mathcal{N}^2$ terms in the sum Eq.~\ref{eq:needforlater2}, but they cannot \emph{all} be the same since 
 \begin{equation}
 [\sigma_I,\sigma_J] \textrm{ is orthogonal to } [\sigma_I,\sigma_{J'}] \textrm{ if } J \neq J'. \label{eq:orthogonalityconditionforPaulicommutators} 
 \end{equation}
 Taking that into account, the worst-case scenario is $n = \mathcal{N}$, which gives the $\sqrt{\mathcal{N}}$ deterioration from the no-interference answer 
 to our actual bound Eq.~\ref{eq:trotterbiinvariantquotesappendix}. 
 
 There are a number of ways we might seek to beat down this scaling.  One way would be to argue that you would have to be fantastically unlucky \cite{Chen2021} for all the signs to conspire such that all the interference is constructive.  Another way would be to argue that given the commutator structure of the generalized Pauli's  it is actually impossible to be so unlucky. 
 
 We'll take a different approach. Rather than rely on luck, we're going to take our fate into our own hands. We're going to use our freedom to re-order the terms in the Trotter product to make sure that the interference is on-net destructive. We will arrange the signs so that interference now works for us, and guarantee that the O($\delta^2$) error term is even smaller than if there were no interference at all. 
 
 In constructing the Trotter product, we could have put the monomial factors  $e^{i h_I \sigma_I \delta}$ in any order. The number of possible orders is $\mathcal{N}!$. The effect of changing the ordering will be to change the signs of some of the terms in the sum in Eq.~\ref{eq:needforlater2}. Changing the sign of a given term will flip between constructive and destructive interference. 
 
Let's write down an iterative procedure for arranging a favorable ordering. We will be handed each component monomial $e^{i h_I \sigma_I \delta}$ in turn, and then we will do one of two things: we will either place the monomial at the very back of the Trotter product, or we will place it at the very front. We will make this choice based on which makes the interference most destructive. Specifically, we will choose to greedily minimize the Frobenius error $||\frac{1}{2} \sum_{I>J} h_I h_J [\sigma_I,\sigma_J]||_{\overline{F}}$, where the sum is taken only over the terms we have already placed in the Trotter product (and not those still waiting to be placed). 

Let's see how this works.  The first two monomials we can place in either order, let's choose $e^{i h_a \sigma_a \delta} e^{i h_b \sigma_b \delta}$. At this stage the O($\delta^2$) error has just a single contribution, $H_{a,b} \equiv \frac{1}{2} h_a h_b [\sigma_a,\sigma_b]$. We are now handed the third monomial. With only three Pauli's in play there's still no interference (because of Eq.~\ref{eq:orthogonalityconditionforPaulicommutators}), so we can place the third monomial anywhere. Let's put it at the back, giving $e^{i h_a \sigma_a \delta} e^{i h_b \sigma_b \delta} e^{i h_c \sigma_c \delta}$. This makes an additional contribution to the O($\delta^2$) error of $H_{a,bc} \equiv \frac{1}{2}  [h_a \sigma_a + h_b \sigma_b,h_c \sigma_c]$, giving 
\begin{equation}
||H_{a,b} + H_{ab,c}||_{\overline{F}}^2 = ||H_{a,b}||_{\overline{F}}^2 + ||H_{ab,c}||_{\overline{F}}^2 \leq h_a^2 h_b^2 + h_a^2 h_c^2 + h_b^2 h_c^2 \ . 
\end{equation} 
It is when we are handed the fourth monomial that things finally get interesting. We will either place the fourth monomial at the very back  $e^{i h_a \sigma_a \delta} e^{i h_b \sigma_b \delta}e^{i h_c \sigma_c \delta} e^{i h_d \sigma_d \delta}$ or at the very front  $e^{i h_d \sigma_d \delta} e^{i h_a \sigma_a \delta} e^{i h_b \sigma_b \delta} e^{i h_c \sigma_c \delta} $. (One could also imagine placing it in the middle, but we won't need to make use of that possibility.) The new contribution to the O($\delta^2$) error will be $H_{abc,d} \equiv  [h_a \sigma_a + h_b \sigma_b + h_c \sigma_c,h_d \sigma_d]$ if we placed $e^{i h_d \sigma_d \delta}$ at the back, or $-H_{abc,d}$  if we placed $e^{i h_d \sigma_d \delta}$ at the front. The norm-squared of the error is 
\begin{equation}
     ||H_{a,b} + H_{ab,c} \pm H_{abc,d} ||_{\overline{F}}^2 
    =       ||H_{a,b} + H_{ab,c} ||_{\overline{F}}^2 + || H_{abc,d} ||_{\overline{F}}^2  \pm 2 \overline{\textrm{Tr}}[(H_{a,b} + H_{ab,c})H_{abc,d} ] \ .  
\end{equation}
If the last term contributes positively we have constructive interference; if it contributes negatively we have destructive interference. We get to choose, and we choose destruction. With this choice we have \begin{equation}
    \textrm{min}_{\pm} \Bigl[ \,      ||H_{a,b} + H_{ab,c} \pm H_{abc,d} ||_{\overline{F}}^2  \, \Bigl] 
    \ \leq \     ||H_{a,b} ||_{\overline{F}}^2 + ||H_{ab,c} ||_{\overline{F}}^2 + || H_{abc,d} ||_{\overline{F}}^2  \ .
    \end{equation}
By iterating this procedure for each monomial in turn---adding it either to the very front or the very back, guided by which gives the most destructive interference---we arrive at 
\begin{equation}
  \textrm{optimal order: } \ \ \ \ \Bigl|\Bigl| \frac{1}{2} \sum_{I= 1}^\mathcal{N} \sum_{J<I}  h_I h_J [ \sigma_I,\sigma_J] \Bigl|\Bigl|_{\overline{F}}^{\ 2}  \ \leq \  { \sum_{I= 1}^\mathcal{N} \sum_{J=1}^{I-1}  h_I^2 h_J^2 } \ \leq \ \frac{1}{2} (\overline{\textrm{Tr}} H^2)^2 \ . \label{eq:optimalorderingresult} 
\end{equation}
Thus by judicious choice of ordering we can make the O($\delta^2$) Frobenius error for the first-order Trotter expansion be parametrically smaller in $\mathcal{N}$ than the worst-case ordering, and indeed can ensure that it is always smaller than if there were no interference at all. 

To make this enhanced bound on the Frobenius error rigorous, we'd need to show that the O($\delta^3$) terms  do not dominate the O($\delta^2$) term. Such an enhanced bound would be particularly intriguing for what it would imply about the diameter of the complexity geometry. Combining the enhanced bound with the method of Sec.~\ref{subsec:implicationsfordiameter} (and folding the approximations of Secs.~\ref{subsec:approx2} and \ref{subsec:trottererror} into a single step) would give an improved lowerbound on the diameter of complexity geometries, and this improved lowerbound would exactly match what we derived by a completely different (differential geometry) technique in \cite{Brown:2021euk}. 

(Note that our sorting procedure does not expend all of our freedom to re-order, since it can only  reach $2^{\mathcal{N}}$ out of the $\mathcal{N}!$ possible orderings. This additional unused freedom might be of assistance in generalizing this method to higher-order  Suzuki-Trotter formulas.)

\section{When does close in complexity geometry imply close in operator norm?} 

Let's prove a further lemma. This lemma was not needed to establish the results of this paper, but may be useful in attempting to improve them. We will derive a sufficient condition on the penalty schedule such that two unitaries being close in the complexity geometry implies that they are also close in the operator norm.

Let's consider the minimal geodesic through the complexity geometry. This will be generated by some  Hamiltonian $H(t) = \sum_I h_I(t) \sigma_I$. The length is 
\begin{equation}
    L = \int dt \sqrt{ \sum_I \mathcal{I}(\sigma_I) h_I(t)^2} \ . 
\end{equation}
On the other hand Eq.~\ref{eq:normpathlessthanhamiltonian} tells us that operator norm is upperbounded by 
\begin{equation}
    ||\mathcal{P} e^{i \int dt H(t)}  - \mathds{1}||_\textrm{op} \ \leq \ \int dt ||H(t) ||_\textrm{op} \ = \int dt || \sum_I h_I(t) \sigma_I ||_\textrm{op} \ \leq \ \int dt \sum_I |h_I(t)|  \ . 
\end{equation}
How large can $\sum_I |h_I|$ be at fixed $\sum_I \mathcal{I}(\sigma_I) h_I^{\, 2}$? For some Lagrange multiplier $c$, the maximum is given by 
\begin{equation}
h_I = \frac{c}{\mathcal{I}(\sigma_I)} \ \ \ \  \rightarrow  \ \ \ \ 
\sum_I \mathcal{I}(\sigma_I) h_I(t)^2 = \sum_I \frac{c^2}{\mathcal{I}(\sigma_I)} \ \    \ \&  \ \ \sum_I |h_I|  = \sum_I \frac{c}{\mathcal{I}(\sigma_I) } \ . 
\end{equation}
Eliminating $c$, and using bi-invariance, gives the bound 
\begin{equation}
\boxed{||U_1 - U_2 ||_\textrm{op} \ \leq \ \sqrt{ \sum_I \frac{1}{\mathcal{I}(\sigma_I)} } \, L(U_1,U_2) }  \ . 
\end{equation} 
This tells us that polynomially small complexity distance implies polynomially small operator-norm distance so long as the penalty schedule is sufficiently punitive,
\begin{equation}
\textrm{sufficient condition: } \ \ \sum_I \frac{1}{\mathcal{I}(\sigma_I)} \textrm{ is at-worst polynomially big} . \label{eq:sufficientconditionoperatoriscomplex}
\end{equation}
This condition is satisfied by the cliff metric Eq.~\ref{eq:examplemetric1} with $ \mathcal{I}_\textrm{cliff} > 4^N$ and the binomial metric Eq.~\ref{eq:examplemetric2} with $\alpha \ \geq 1$. 
As discussed in Sec.~\ref{subsec:vastrightinvariantconspiracy}, this is yet another place in which the $\alpha = 1$ binomial schedule shows up as a critical metric.  

We can also \emph{lower}bound the operator-norm distance. Combining Eqs.~\ref{eq:tobeprovedinappendix1}, \ref{eq:norminequalitieswith2totheN}, and  \ref{eq:conservedhardness}  gives $L/\Gamma  \leq s  \leq \frac{\pi}{2}||U_1-U_2||_{\overline{F}} \leq \frac{\pi}{2}||U_1-U_2||_\textrm{op}$ and therefore  
\begin{equation}
\frac{2}{\pi} \sqrt{ \frac{1}{\mathcal{I}_{\textrm{max}}} } \, L(U_1,U_2) \ \leq \ ||U_1 - U_2 ||_\textrm{op} \ \leq \ \sqrt{ \sum_I \frac{1}{\mathcal{I}(\sigma_I)} } \, L(U_1,U_2)   \ . 
\end{equation} 
The lowerbound is saturated by those monomial unitaries $e^{i \sigma_I \delta}$ that are generated by a short step in the most expensive generalized-Pauli directions, $\mathcal{I}(\sigma_I) = \mathcal{I}_\textrm{max}$. 


\end{document}